\documentclass[prb,twocolumn,superscriptaddress,showpacs]{revtex4}
\usepackage{epsfig}
\usepackage{latexsym} 
\usepackage{bm}
\newcommand{\fig}[2]{\includegraphics[width=#1]{#2}}

\begin{document}

\newcommand{\be}{\begin{equation}}
\newcommand{\ee}{\end{equation}}
\newcommand{\bea}{\begin{eqnarray}}
\newcommand{\eea}{\end{eqnarray}}
\newcommand{\nn}{\nonumber}

\newcommand{\dg}{{\dagger}}
\newcommand{\pdg}{{\vphantom\dagger}}

\newcommand{\ra}{{\rangle}}
\newcommand{\la}{{\langle}}

\newcommand{\br}{{\bf{r}}}
\newcommand{\hT}{{\hat{T}}}
\newcommand{\cU}{{\cal U}}
\newcommand{\cD}{{\cal D}}
\newcommand{\cA}{{\cal A}}
\newcommand{\cJ}{{\cal J}}

\newcommand{\cis}{{\cal I}^\ast}
\newcommand{\cise}{{\cal I}^\ast_{even}}
\newcommand{\ciso}{{\cal I}^\ast_{odd}}
\newcommand{\sfs}{{\cal SF}^\ast}
\newcommand{\supf}{{\cal SF}}
\newcommand{\sfse}{{\cal SF}^\ast_{even}}
\newcommand{\sfso}{{\cal SF}^\ast_{odd}}
\newcommand{\fls}{{\cal FL}^\ast}
\newcommand{\flse}{{\cal FL}^\ast_{even}}
\newcommand{\flso}{{\cal FL}^\ast_{odd}}

\newcommand{\bk}{{\bf k}}
\newcommand{\bq}{{\bf q}}

%%%%%%%%%%%%%%%%%%%%%%%%%%%%%%%%%%%%%%%%%%%%%%%%%%%%%%%%%%%%%%%%%%%%%%

\title{Extending Luttinger's theorem to
$Z_2$ fractionalized phases of matter}

\date{\today}

\author{Arun Paramekanti}
\affiliation{Department of Physics and Kavli Institute for
Theoretical Physics, University of California, Santa Barbara
CA 93106-4030}
\author{Ashvin Vishwanath}
\affiliation{Department of Physics, Massachussetts Institute of
Technology, 77 Massachussetts Ave., Cambridge MA 02139}
\begin{abstract}
\vspace{0.1cm}
Luttinger's theorem for Fermi liquids equates the volume enclosed by the Fermi
surface in momentum space to the electron filling, independent of the strength and
nature of interactions. Motivated by recent momentum balance arguments that
establish this result in a non-perturbative fashion [M. Oshikawa, Phys. Rev. Lett. {\bf
84}, 3370 (2000)], we present extensions of this momentum balance argument to exotic
systems which exhibit quantum number fractionalization focussing on $Z_2$
fractionalized insulators, superfluids and Fermi liquids. These lead to nontrivial
relations between the particle filling and some intrinsic property of these quantum phases,
and hence may be regarded as natural extensions of Luttinger's theorem. We find that
there is an important distinction between fractionalized states arising naturally
from half filling versus those arising from integer filling.  We also note how these
results can be useful for identifying fractionalized states in numerical
experiments.
\typeout{polishabstract} 
\end{abstract}

\pacs{71.27.+a,71.10.-w,75.10.Jm,75.40.Mg}
\maketitle

\section{introduction}

The last two decades have witnessed the experimental discovery of several strongly
correlated materials that show properties strikingly different from that expected from
conventional theories based on Landau's Fermi liquid picture. These include the high
temperature copper oxide superconductors, heavy fermion systems near a quantum
critical point, and, more recently, the cobalt oxide materials. Interesting
correlated quantum phases are also likely to emerge in the near future from ongoing
experimental efforts in the area of cold atoms in optical lattices. It has then become
imperative to theoretically investigate quantum phases of matter that differ
fundamentally from the standard paradigm. Indeed, in such a search for new
theoretical models, it would be useful to know if general principles place
constraints on the possible quantum phases. Here, we will explore in detail the
consequences of one such constraint, arising from momentum balance, which will be
applicable to interacting many body systems on a lattice. This argument was 
first
applied to the case of one-dimensional Luttinger liquids \cite{Oshikawa97},
where it relates
the Fermi wavevector $k_F$ to the particle density. It was later extended
to Fermi liquids \cite{Oshikawa00} in dimensions $D \geq 2$,
where it leads to
Luttinger's theorem \cite{Luttinger60}, relating the filling fraction to the volume
enclosed within the Fermi surface on which the long lived Fermi-liquid
quasiparticles are defined.
Here, we will apply the same line of argument to a
variety of different phases in spatial dimensions $D>1$, and the constraint we
obtain in this way may be viewed as analogues of Luttinger's theorem for these
phases. 
In all cases, the filling fraction (number of particles per unit cell of the
lattice) is fundamentally related to some intrinsic property of the phase. 

The momentum balance argument, introduced by Oshikawa for Fermi Liquids
\cite{Oshikawa00}, proceeds as
follows. Consider a system of interacting fermions at some particular filling on a
finite lattice at zero temperature. Periodic boundary conditions implies that the
lattice has a torus geometry; imagine introducing a solenoid of flux in one of the
holes of the torus, and adiabatically changing its strength from zero to $2\pi$. The
crystal momentum imparted to the system can then be caculated in two different ways.
First, in a trivial fashion that only depends on the filling and is independent of
the quantum phase the system reaches in the thermodynamic limit, and second, in way
that depends essentially on the quantum phase of the system. Consistency requires
the equality of the these two quantities - which leads to the nontrivial conditions
on the quantum phase. Essentially, each consistent quantum phase has its own way of
absorbing the filling dependent crytal momentum that is generated in this process -
as mentioned, 
in the case of the Fermi liquid this leads to Luttinger's theorem.

Here, we begin by applying this argument to the case of (bosonic) insulators at half
filling, where the system in the thermodynamic limit
necessarily acquires an enlarged unit cell (through broken translational
symmetry or a spontaneous flux)
or develops
topological order. For the latter case, the momentum balance argument fixes the
crystal momentum of the degenerate ground states in the different topological
sectors. A useful side result of this analysis will be a general prescription to
distinguish between between a $Z_2$ fractionalized insulator (or spin liquid) and a
more conventional translation symmetry broken state, which is useful when the the
order parameter for the translation symmetry breaking is not obvious. This is
relevant for numerical studies on finite sized spin systems that search for
fractionalized spin liquid states \cite{Misguich_numerics,Sheng04}. 

Next, we apply the same methods to the case of exotic fermi liquids ($FL^*$ phases)
proposed recently in Ref.~\onlinecite{Senthil_FLstar}. This is a phase which has
conventional electron like excitations near a Fermi surface, but also posesses
topological order and gapped fractionalized excitations. The question of interest
here is whether the Fermi surface in these systems `violate' Luttinger's theorem
(given that these phases are not continuously connected to the free electron gas this
is of course not prohibited by Luttinger's proof\cite{Luttinger60}), 
and if so whether there is a
generalization of Luttinger's theorem that can accomodate these cases as well.
Indeed, applying the momentum balance argument to the $FL^*$ phases we find that
while Luttinger's Theorem is violated by these Fermi volumes, this violation is not
arbitrary but is constrained to be one of a few possibilities which is determined uniquely by
the pattern of fractionalization. 

Finally, we apply these arguments to the case of neutral superfluids. For
conventional superfluids we argue that
this leads to a constraint on the Berry phase acquired on
adiabatically moving a vortex around a closed loop, by relating it to the boson filling. Loosely
speaking this is the quantity that determined the Magnus `force' on a moving vortex.
Since this relation between the Magnus force and the boson filling is obtained using
the same momentum balance argument that leads to Luttinger's theorem when applied to
a Fermi liquid, it may be viewed as a `Luttinger' theorem for superfluids.
Alternatively, since a similar relation between boson density and Magnus force is
known for superfluids with Galilean invariance \cite{HaldaneWu85,Ao93}, this may be
viewed as an extension of those results to the case of lattice systems. In contrast,
while Galilean invariance also constrains the zero temperature value of the
superfluid stiffness, that constraint does not survive the inclusion of the lattice.
In order to further bring out the similarity of this relation in superluids to the
Luttinger relation, we consider fractionalized superfluid phases $SF^*$, (related to
the exotic superconductor $SC^*$ studied in ref.~\cite{Senthil_Z2}), or equivalently
superfluid analogues of the fractionalized Fermi liquid phases. We show that they
too violate the conventional relation between vortex Berry phase and boson filling
in exactly the same way that Luttinger's theorem is violated in $FL^*$.
We discuss caveats in the relation between the vortex Berry phase and the boson
filling in conventional and fractionalized superfluids which make the above relation less
rigorous at the present time than the analogous relation for Fermi liquids and
insulators.

The relation between vortex Berry phase and boson
filling in lattice superconductors can lead to surprising conclusions in some cases.
For example, consider a conventional superfluid (where the bosons are Cooper pairs
of electrons) obtained by doping a band insulator versus another conventional
superfluid obtained on doping a proximate Mott insulator. One may imagine that only
the doped charges participate in the superfluidity - indeed this is roughly what is
expected for a quantity like the superfluid stiffness (although it is not strictly
constrained in these lattice systems\cite{paramekanti98}). 
However, a result of the discussion below
will be that the Berry phase acquired by a vortex in this system arises from
counting {\em all} charges in the system (and not just the charges doped into the
Mott insulator). In this sense at zero temperature all particles participate in the
superfluidity. In contrast, an exotic superfluid phase $SF^*$ 
can display a phase
where only the doped charges contribute to the Berry phase. 

A recurring theme throughout this paper will be the distinction between exotic
phases obtained from a correlated `band' insulator (i.e. one that has an interger
filling per site in the case of bosons) and which could form a translationally
invariant conve ntional insulating state;  versus those obtained by from an exotic
phase at half filling. For example, if we consider featureless $Z_2$ fractionalized
insulators at integer and half integer filling, then at low energies they are be
described by `even' and `odd' $Z_2$ gauge theories respectively (in the terminology
of Ref.~\onlinecite{Moessner02}).  The different ground state
topological sectors of the odd gauge theory
can in certain geometries carry a finite crystal momentum, while the the crystal
momenta associated with the different ground state
sectors of an even gauge theory are always
zero. This distinction persists if these phases are doped to obtain exotic Fermi
liquids and superfluids.  The distinction is especially striking in the case of
$FL^*_{odd}$ where a Fermi volume that violates Luttinger's theorem arises.  In
contrast, $FL^*_{even}$ obeys Luttinger's theorem but is nevertheless an exotic
phase. A similar distinction will apply to the exotic superfluids - in that case the
relation between the Magnus `force' on a vortex and the filling is the regular one
for $SF^*_{even}$ but is unconventional in the case of $SF^*_{odd}$. 

An essential ingredient in the following arguments will be the evolution of a
quantum state under flux insertion. While this recalls the argument of Laughlin
\cite{Laughlin81} for the integer Quantum Hall effect, there is an important
distinction that must be noted. In the case of Laughlin's argument and a similar
argument for the forces on superfluid vortices given by Wexler\cite{Wexler97}, the
conclusions are derived by keeping track of the change in energy during the process
of flux threading. More recently, rigorous energy counting arguments for charge and
spin insulators have been made by Oshikawa \cite{Oshikawa03} and Hastings 
\cite{Hastings}.
In contrast here we will follow Ref.~\onlinecite{Oshikawa00} and rather
keep track of the change in {\em crystal momentum} during the flux threading
process, which will allow us to derive 
a different set of rather general conclusions that apply to a
variety of phases. We also note that while the subject of Magnus force on a vortex
at finite temperatures, in the presence of quasiparticle or superfluid phonon
excitations, has been the subject of much lively debate (see for example
Refs.~\onlinecite{controversy}), our arguments will only apply to the case of zero temperature
and hence cannot address any of the issues under debate.

The layout of this paper is as follows. Due to the length of the paper we give in
Section \ref{overview} a simplified overview of all results and a brief
indication of the method used. Then, we pass to the technical details and in section
\ref{counting} review the momentum counting procedure which will be applied to all
the phases.  Then, we consider conventional insulators in section \ref{conventionalI}
and $Z_2$ fractionalized insulators in section \ref{Z2insulator} using the
momentum balance argument and discuss how they may be unambiguously distinguished in
numerical experiments in section \ref{numerics}.  We then review the momentum
balance argument \cite{Oshikawa00} for conventional Fermi liquids in section
\ref{conventionalFL}, which leads to Luttinger's theorem, and see how this is
modified ina systematic way when applied to $Z_2$ fractionalized Fermi liquids in
section \ref{FLstar}. Next we apply these arguments to conventional, neutral
superfluids and $Z_2$ fractionalized superflui ds, in Sections \ref{conventionalSF}
and \ref{SFstar} respectively. We conclude with some observable consequences that
arise directly from these considerations.

\section{Overview of the Momentum Balance Argument}
\label{overview}

In this section we summarize the results of the momentum balance argument applied to
different phases. While the detailed arguments leading to these results are
contained in the following sections, the results themselves are easily stated, which
is done below along with some heuristic supporting arguments. 

\begin{figure}
\begin{center}
\vskip2mm
\hspace*{0mm}
\centerline{\fig{1.5in}{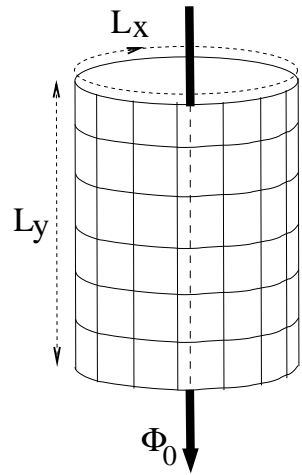}}
\vskip2mm
\caption{Schematic figure showing flux threading in a cylinder
geometry, with flux $\Phi_0=hc/Q$.}
\label{cylinder}
\end{center}
\end{figure}

Consider the system in a cylindrical geometry, as shown in Fig.~\ref{cylinder} with
dimensions $L_x$, $L_y$ (integers), and a total of $N$ particles (bosons or spinless
fermions). Now imagine adiabatically threading a flux of $2\pi$ through the center
of the cylinder - the particles are assumed to couple minimally to this flux with a
unit charge The total momentum imparted to the system can be calculated using
Faraday's law $F_x = -\frac1{L_x}\frac{d\phi}{dt}$ and integrating this force over
time leads to the change in momentum 
\be
\Delta P_x = \frac{2\pi}{L_x} N, 
\label{trivialpx}
\ee
which is
only defined modulo $2\pi$ since the system is on a lattice. A more rigorous
calculation in the following sections arrives at the same result. A similar
procedure performed along the perpendicular direction (for which it is convenient to
think of the system living on a torus) yields the other component of the crystal
momentum 
\be
\Delta P_y = \frac{2\pi}{L_y} N.
\label{trivialpy}
\ee
This is the result of trivial momentum
counting - we now consider how this additional crystal momentum is accomodated in
the various different phases. 
For later purposes, we will define the ``filling'' $\nu \equiv N/L_x L_y$,
which is the number of particles per unit cell.
%%%%%%%%%%%%%%%%%%%%%%%%%%%%%%%%%%%%%%%%%%%%%%%%%%%%%%%%%%%%%%%%%%%%%%
%%%%%%%%%%%%%%%%%%%%%%%%%%%%%%%%%%%%%%%%%%%%%%%%%%%%%%%%%%%%%%%%%%%%%%

\noindent{\bf Insulators:} Consider the evolution of the ground state of a many body
system under the adiabatic insertion of a $2\pi$ flux. For an insulator, it may be
easily verified that the final state must have an energy equal to that of the ground
state, in the thermodynamic limit. That is, the system ends up either in the
original state, or in a degenerate ground state. This is just because the change in
energy under flux threading is related to the average current 
$\langle J \rangle =
\frac{dE}{d\phi}$. Since in the insulating state the current must vanish in 
the
thermodynamic limit, the energy of the final state must equal that of 
the initial state. 

If the insulator is at integer filling, i.e. $N= \nu L_x L_y$, with the filling
$\nu$ such that $\nu \in \mbox{integer}$, then we have ($\Delta P_x = 2\pi \nu L_y$,
$\Delta P_y = 2\pi \nu L_x$), and so $\Delta P_x \equiv \Delta P_y \equiv 0
(\mbox{mod } 2\pi)$. This is compatible with the system having a unique ground state
which it returns to at the end of the $2\pi$ flux threading. This is the
conventional featureless insulating state (band insulator for Fermions, integer
filling Mott insulator for bosons) - although in principle more complicated states
are possible at integer filling as well. 

The case of insulators at non-integer filling is more interesting. For definiteness,
consider bosons at half filling ($\nu = 1/2$). In order that the total number of
particles be an integer, we need the total number of sites $L_x \times L_y$ to be
even. If we first consider the case $L_y$ odd and $L_x$ even, under flux insertion
in the geometry of Fig.~\ref{cylinder} we will have $\Delta P_x = \pi (\mbox{mod}
2\pi)$. Therefore, the initial and final states (which we have argued to be
denenerate in the thermodynamic limit) must differ in crystal momentum and hence one
is forced to conclude that the ground state is at least doubly degenerate in this
even$\times$odd geometry. Such a degeneracy can result from one of two different
reasons (we assume that time reversal symmetry is not spontaneously
broken and the special case of $\pi$-flux is discussed in Section~\ref{conventionalI} D). 
First, the system may be heading towards translation symmetry breaking in
the thermodynamic limit. In this case we can form the symmetric and antisymmetric
combinations of the two ground states which clearly transform into each other under
a unit horizontal translation. Translation symmetry breaking implies that there is
a {\it local} operator that can distinguish between these two states. For example,
if the system is heading towards a CDW state, (eg. with a stripe pattern with the
charge on alternating columns in Fig.~\ref{cylinder}), then the relevant local
operator is simply the charge density, which would distinguish these two states as
being translated versions of one another. It may of course happen that the relevant
local operator is less obvious (eg. bond centered charge density) but nevertheless
in principle this distinction between the two states can be made with some local
operator. It may however happen that there exists no local operator that can
distinguish these two states. Then, the system will appear perfectly translation
symmetric, although it is an insulator at half filling. Indeed this is precisely the
property of the RVB spin liquid state proposed for spin $1/2$ lattice systems with
one spin per unit cell - which can also be cast in the language of the above
discussion on identifying the spins with hard core bosons. Therefore, these
degenerate ground states can only be distinguished via a non-local operator. This is
called topological order, where degeneracies arise on spaces with non trivial
topology that are related to creating a topological excitation which is highly
nonlocal in the original variables. In the following sections we consider one
concrete theoretical realization of this scenario - the case of a deconfined $Z_2$
gauge theory coupled to bosons carrying one half the elementary unit of charge. The
translationally symmetric state at half filling may be roughly visualised as a
uniform state with a half charge at each lattice site. The degenerate ground states
correspond to the topological degeneracy of the theory on a cylinder - which is
relate to the presence or absence of an Ising flux (vison) in the hole of a
cylinder. In the following sections we explicitly demonstrate that threading a
$2\pi$ flux induces such a topological excitation and causes the ground state to
evolve into this distinct topological sector. Thus, the way these topologically
ordered states accomodate the crystal momentum imparted to the system on flux
threading, despite being translationally symmetric, is by creating a vison
excitation in the hole of the cylinder which then carries the appropriate crystal
momentum. Later we will draw a distinction between Ising gauge theories where the
vison carries a crystal momentum (odd gauge theories) and those where it does not
carry momentum (even gauge theories). 

% For translationally symmetric $Z_2$ insulators, this will be 
%determined by whether the filling is half-odd-integer or integer. 

\noindent{\bf Fermi Liquids:} The original application of the momentum 
balance
argument was to the case of Fermi liquids in Ref.~\onlinecite{Oshikawa00}. This
argument is reviewed in the following sections - here we just note the
main points. If we begin with spinless electrons at a filling $\nu$,
then the flux threading excites quasiparticles around the Fermi
surface. The total crystal momentum carried by these excitations can
be converted into an integral over the volume enclosed by the Fermi
surface, which leads to the relation between the filling (which enters
the trivial momentum counting) and the Fermi volume. For an
appropriately chosen system size, these lead to the relation:
\begin{equation}
\nu = \frac{V_{FS}}{(2\pi)^2} + p
\label{lutt}
\end{equation}
where $V_{FS}$ is the Fermi volume, and $p$ is an arbitrary integer
which represents the filled bands. This is just Luttinger's Theorem -
in Ref.~\onlinecite{Oshikawa00} it was also applied to Kondo lattice models where
it yields the large Fermi surface expected in the Kondo screened
phase.

One can now ask the reverse question --- given a phase whose low energy
exciations are electron like Landau quasiparticles, does this phase
necessarily also  satisfy Luttinger's theorem? From the above
momentum balance argument it is clear that in order to violate the
Luttinger relation there must exist an alternate sink for the
momentum. From our previous discussion of topologically ordered states,
it is clear that if topological order coexists with Fermi liquid
like excitations, then the momentum balance can be satisfied with a
non-Luttinger Fermi volume. In fact the specific case of an exotic
Fermi liquid with $Z_2$ topological order ($FL^*$ phase) was proposed
in Ref.~\onlinecite{Senthil_FLstar} in the context of the heavy fermion
systems. This phase has low energy excitations identical to that of a
Landau Fermi liquid of electrons, but also a gapped Ising vortex
excitation. The Luttinger relation relating the filling to the volume
enclosed by the Fermi surface of these quasiparticles can then be
violated, but in a very specific way. Flux threading creates a Ising
vortex which carries crystal momentum $\pi$ (in an odd$\times$even
system), while the remaining momentum is absorbed by the quasiparticle
excitations. This leads to the modified Luttinger relation:

\begin{equation}
\nu = \frac12 + \frac{V_{FS}}{(2\pi)^2} + p
\end{equation}
where $p$ is an arbitrary integer representing filled bands. Note the
crucial difference from Luttinger's relation in Eqn.\ref{lutt}, that
arises from the extra factor of $\frac12$. Clearly, this is related to
the fact that a translationally invariant insulator is possible at
half filling, where the Fermi volume can shrink to zero. Thus, the
difference from the original Luttinger relation is very specific,
i.e. the filling of exactly half a band, for the case of $Z_2$
fractionalization. We have commented earlier on the difference between
`odd' vs. `even' $Z_2$ gauge theories, again this distinction is
crucial here and moreover is not directly set by the filling as it was
for the case of the translationally symmetric insulating states. The
above violation of the Luttinger relation only occurs in the case of
the `odd' gauge theory, $FL^*_{odd}$.

\noindent{\bf Superfluids:} Finally, we consider the case of neutral
superfluids. Here threading a $2\pi$ flux clearly inserts a vortex
through the hole of the cylinder. This can also be visualized as
creating a vortex in the superfluid at the bottom of the cylinder and
dragging it all the way to the top. Clearly such a  vortex will
experience a `Magnus force' in the direction perpendicular to its
motion. Let us ignore for a moment the lattice and calculate the
momentum imparted by this force $F_M = 2\pi\alpha_M \vec{v}\times
\hat{z}$, where $\vec{v}$ is the velocity and $\alpha_M$ a constant
that fixes the Magnus force. The total momentum transferred to the
system is then independent of the details of the vortex motion and
depends only on its net displacement --- this yields $\Delta P_x = 2\pi
\alpha_M L_y$. Equating this to the momentum obtained from trivial
momentum counting (\ref{trivialpx}) and reintroducing the lattice heuristically
by allowing the
the momentum to change in arbitrary integer multiples of $2\pi$ we have:

\begin{equation}
\nu = \alpha_M + p
\label{luttsf}
\end{equation}
where $p$ is an arbitrary integer. Thus, the fractional part of
$\alpha_M$ is completely determined by the boson filling $\nu$. This
can be viewed as the analogue of Luttinger's theorem for bosonic
system, since it is obtained using the same line of argument. It can
also be viewed as an extension of the well known equivalent result for
Galilean invariant superfluids \cite{HaldaneWu85}
(where the Magnus coefficient is the
boson density) to the case of lattice superfluids. While we have been
interpreting $\alpha_M$ above in terms of the `Magnus force' clearly
this is not a well defined concept in a lattice system. In fact, the
property that is sharply fixed by $\alpha_M$ is the Berry phase
acquired by a vortex on adiabatically taking it around a big loop of size
$\mathcal{N}$ plaquettes, which will be shown to be  $2\pi\alpha_M
\mathcal{N}$. This is related to the well known relation \cite{HaldaneWu85} 
in Galilean superfluids between the Magnus force and the Berry
phase acquired by a vortex. 

Again, one can ask if the relation is Eqn.(\ref{luttsf}) can be
violated in any kind of superfluid. Indeed, topologically ordered
superfluid states $SF^*$ can be defined in complete analogy with
$FL^*$. For the particular case of an exotic superfluid state $SF^*$
with $Z_2$ topological order \cite{Senthil_Z2}, there exists in
addition to the usual vortex excitation, an Ising vortex excitation as
well. Threading $2\pi$ flux then creates both a regular vortex and an
Ising vortex - the latter can carry crystal momentum $\pi$ (in the
phase $SF^*_{odd}$ - in which case the remaining momentum is
associated with the `Magnus force' on the superfluid vortex. Then, in
this case as well, the Magnus coefficient $\alpha_M$, associated with
the vortex Berry phase satisfies:

\begin{equation}
\nu = \frac12 + \alpha_M^* + p
\end{equation}
where $p$ is an arbitrary integer. Thus, the `Luttinger relation' for
a conventional superfluid in Eqn.(\ref{luttsf}) is violated, in
exactly the same way that $FL^*_{odd}$ violates the Luttinger relation
for conventional Fermi liquids.

\section{Trivial Momentum Counting}
\label{counting}

Consider a system of $N$ particles, each with charge $Q$, living
on an $L_x\times L_y$ lattice wrapped into the form of a torus 
with periodic boundary conditions along both directions. 
The main result of this section is
that if one adiabatically threads flux $hc/Q$ through one of
the holes of the
torus, the crystal momentum difference between the initial and final 
state is:
\be
P_f - P_i = 2\pi N/L_x (\mbox{mod }2\pi) = 2\pi\nu L_y (\mbox{mod }2\pi)
\ee
where $\nu = N/(L_x L_y)$ is the charge density in units of $Q$
(the ``filling'').
This result is independent of the eventual quantum phase of the
system in the thermodynamic limit, and we will refer to it as the
``trivial'' counting.
Although this has been shown in Ref.~\onlinecite{Oshikawa00}, we include
a derivation here for the sake of completeness, and to fix notation.

The Hamiltonian for an interacting set
of particles (fermions or bosons) coupled to an external vector potential 
can be written down as $H_A = \hat{K}_A + V[\hat{n}]$, with the kinetic 
energy
$\hat{K}_A$, in the presence of a vector potential $A_{ij}$, given by
\be
\hat{K}_A = -\frac{1}{2} \sum_{ij} \left[ t_{ij} e^{-i Q A_{ij}/\hbar
c} B^\dg_i B^\pdg_j + h.c.  \right]
\label{Hgeneral}
\ee
$B^\dg_i$ creates a particle carrying charge-$Q$ at
site $i$.  The interaction term $V[\hat{n}]$ depends only on the
density of the particles. Both $t_{ij}$ and $V[\hat{n}]$ are 
invariant under lattice translations.
For simplicity of presentation, we will assume that 
$t_{ij}$ only connects nearest neighbor sites on a square lattice,
with unit lattice spacing.

To thread a unit flux $\Phi_0=hc/Q$ through the hole of the torus, say along
the $-\hat{y}$ axis as in Fig.~\ref{cylinder}, 
we can choose a uniform gauge in which
$A_{i,i+\hat{x}}=-\Phi(t)/L_x$ and $A_{ij}=0$ for other links, and
adiabatically increase $\Phi(t): 0\to hc/Q$. The state
we reach on flux insertion can of course be written
as $\vert \Psi(T)\ra = \cU_T
\vert\Psi(0)\ra$ with the unitary time evolution operator $\cU_T =
{\cal T}_t \exp\left(-i \int_0^T H_A(t) dt\right)$ where ${\cal T}_t$
is the time-ordering operator. In the final Hamiltonian, the vector
potential corresponds to flux $\Phi(T)=\Phi_0$.

Clearly, the initial and final wavefunctions, as well as the Hamiltonian,
transform under gauge transformations.
Thus, since the final Hamiltonian includes a unit flux quantum,
we need to fix a gauge in order to consistently
define the crystal momentum of a state as the eigenvalue of the unit
lattice translation operator acting on the state and to compare it for
the two states. We pick a gauge such
that $A_{ij}=0$ in the initial as well as the final Hamiltonian ---
this Hamiltonian is denoted as $H_0$.  In this case, for a
threaded flux $\Phi_0$, we need to make a unitary gauge transformation
\be
H_A(T) \to \cU_G H_A(T) \cU^{-1}_G = H_0
\ee
with the operator 
\be
\cU_G = \exp( i \frac{2\pi}{L_x} \sum_i x_i \hat{n}_i). 
\ee
The final wavefunction in this gauge is, in obvious notation,
$\vert \Psi_f \ra = \cU_G \cU_T \vert \Psi_i\ra$. To compute the
crystal momentum of this state, we must act on it with the unit
translation operator $\hT$. This defines the initial and final crystal
momenta, $P_i, P_f$ through $\hT \vert \Psi_i \ra = \exp(-i P_i) \vert
\Psi_i \ra$ and $\hT \vert \Psi_f \ra = \exp(-i P_f) \vert \Psi_f
\ra$. Translating the final state we find
\bea
\hT \cU_G \cU_T \vert \Psi_i\ra &=& (\hT \cU_G \hT^{-1}) (\hT \cU_T
\hT^{-1}) \hT \vert \Psi_i\ra \nn \\
&=& (\hT \cU_G \hT^{-1}) \cU_T e^{- i P_i} \vert \Psi_i \ra,
\eea
since the operator $\cU_T$ commutes with the $\hT$ as the
time-dependent
Hamiltonian is translationally invariant in the uniform gauge. At the
same time, it is straightforward to show that
\be
\hT \cU_G \hT^{-1} = \exp(-i 2\pi N/L_x) \cU_G.
\ee
It is then clear that $P_f = P_i + 2\pi N/L_x (\mbox{mod }2\pi)$, or defining
the filling $\nu = N/(L_x L_y)$, the change in crystal
momentum is $P_f-P_i=\Delta P = 2\pi\nu L_y (\mbox{mod }2\pi)$.  

It is essential for this argument to go through that one has a conserved
$U(1)$ charge, this permits us to couple the charge to an inserted 
solenoidal flux. One can easily generalize the argument to cases where 
the charged
particles carry spin and are coupled to spins fixed to the
lattice such as in a Kondo lattice model \cite{Oshikawa00}.
In this case, one can thread a flux which 
couples to a single component of the spin of the charged carriers,
and eliminate the vector potential using a unitary transformation
which acts on the charged particles as well as the fixed spins.

As mentioned earlier, the result above has been derived without any 
assumption about the
thermodynamic phase of the system. Such an assumption is important for
counting the momentum in a second independent way, which provides
constraints on the various quantum phases of the system and we turn to 
this in the remaining Sections.
For convenience of notation, we will set $\hbar\!=\!c\!=\!1$ in most places.

\section{conventional insulators} 
\label{conventionalI}
\subsection{No Broken Symmetry}

Consider a conventional insulator with a unique ground state and a
nonzero gap to current carrying excitations. Under adiabatic flux
threading, since
the Hamiltonian is time-dependent, the rate of change of energy is
given by $\la d\hat{H}/dt \ra = - \sum_{ij} \la \hat{J}_{ij}(t) \ra
\partial A_{ij}(t)/\partial t$ where the current operator
$\hat{J}_{ij} = -i Q t_{ij} (B^\dg_i B^\pdg_j e^{- i A_{ij}(t)} -
h.c.)$ for the Hamiltonian with kinetic energy as in \ref{Hgeneral}.
Let us assume a linear rate of change of $A_{ij}(t)$ (for
$j=i+\hat{x}$) over a time interval $T$ for threading one flux quantum
$\Phi_0=2\pi$, i.e. the electric field $E_{ij}=-\partial A_{ij}/ \partial t
= (2 \pi/ Q L_x T) \hat{x}$ is a constant over the interval $T$.
The total change in
energy is thus 
\be
\delta E = \int_0^T dt \la d\hat{H}/dt \ra = 2
\pi \hbar \bar{I},
\ee
where the average current in units of $Q$ is 
\be
\bar{I} = 1/(\hbar Q L_x T)
\sum_{i}
\int_0^T dt \la \hat{J}_{i,i+\hat{x}}(t) \ra.
\ee
Clearly $\bar{I}=0$ in an
insulator in the thermodynamic limit \cite{footnote.drude}
--- there is no current flow, and
thus $\delta E=0$! 

However, if we thread one flux quantum
into the system it can be eliminated using a gauge transformation
which leaves the spectrum invariant, as is well-known and was shown in
the previous Section. 
Since the system has a unique ground state with
a charge gap, and $\delta E=0$, this means 
the final state and the initial state in the $A_{ij}=0$ gauge must be the
same. Clearly, there is no change in crystal momentum on threading 
flux $\Phi_0$, which implies
\be 
2\pi \nu L_y = 0 (\mbox{mod } 2\pi)
\ee 
for any $L_y$. This is only possibly if
$\nu$ is an integer.  Thus we arrive at the result: {
%\em 
a {\em conventional} insulator with a unique ground state (i.e., no
broken symmetry) and a nonzero gap to charged excitations is only
possible at integer filling.}\cite{Oshikawa00b,Shankar}

\subsection{Conventional insulator with broken translational symmetry}

Imagine tuning the interaction $V[\hat{n}]$ in the above Hamiltonian
in Eq.~\ref{Hgeneral}, such that the ground state of the system in the
thermodynamic limit is an insulator which breaks translational
symmetry.  The thermodynamic ground state is clearly degenerate, the
degeneracy reflecting the different broken symmetry patterns.  On a
finite lattice, such a system must thus have eigenstates with
different crystal momentum, which, in the thermodynamic limit, become
degenerate and allow us to construct linear superposition eigenstates
which break the translational symmetry.

Let us consider such a system on a finite lattice, with aspect ratio
such that the thermodynamic broken symmetry
pattern is not frustrated. If the insulator is stabilized
at a density $\nu=p/q$ (with $p,q$ having no common factors), the flux
threading argument implies the ground state must
evolve under $hc/Q$ flux insertion into a different state which has a 
relative crystal
momentum $\Delta P = 2\pi (p/q) L_y$, with an energy equal to the
ground state energy in the thermodynamic limit. 
These states would be ``quasi-degenerate'' on a large finite 
lattice.\cite{footnote1}

%Let us discuss this, for simplicity, in a system with just 
%two such states on a finite torus. $hc/Q$ flux insertion (in the $A_{ij}=0$
%gauge) in any one hole of the torus
%can then either lead to an interchange of the two states or to
%each state evolving into itself. In both cases, on threading flux $2hc/Q$
%through the same hole (say along $\hat{y}$), the state evolves into itself 
%with a momentum change $4\pi\nu L_y$.
%Thus $4\pi\nu L_y = 0 (\mbox{mod } 2\pi)$, which leads to at most two
%possibilities: $2\pi\nu L_y = 0,\pi$.
%
%Since $L_y$ is an integer, the existence of only these two possibilities
%means $2\nu$ must be an integer. More generally, a conventional insulator 
%with $q$-fold 
%broken translation symmetry on a torus can only be realized at fillings
%$\nu$, such that $q\nu$ is an integer.

\subsection{Flux threading in the conventional broken symmetry insulator}

The manner in which the set of quasi-degenerate 
states in a broken symmetry insulator evolve under adiabatic flux 
insertion is fixed by momentum
balance. Let us again work with a system with a two-fold broken
symmetry in the thermodynamic limit.

\begin{figure}
\begin{center}
\vskip2mm
\hspace*{0mm}
\centerline{\fig{3.5in}{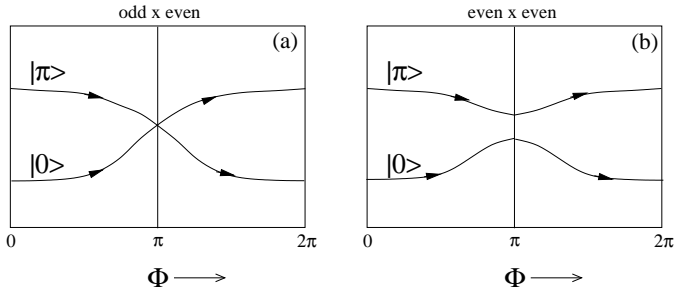}}
\vskip2mm
\caption{Evolution of energy levels upon flux threading in a conventional 
insulator with
two-fold broken translational symmetry on a cylinder.
The two levels which
become degenerate ground states in the thermodynamic limit carry 
momenta $0,\pi$. (a) The two levels cross upon threading flux along
$\hat{y}$ in a geometry with $L_x=$even, $L_y=$odd.
(b) The two levels return to themselves upon threading flux along
$\hat{y}$ in a geometry with $L_x=$even, $L_y=$even.}
\label{crossing}
\end{center}
\end{figure}

If the filling $\nu$ and $L_y$ are such that $2\pi \nu L_y = \pi 
(\mbox{mod } 2\pi)$, flux
insertion causes a momentum change of $\Delta P_x = \pi$. 
This implies we must have
two quasi-degenerate states
differ in $\hat{x}$-crystal momentum by $\pi$, and flux threading 
must lead to an
interchange of these two states. This is depicted schematically 
in Fig.~\ref{crossing}(a) where the
two states on a finite size system, denoted by $\vert 0\ra, \vert \pi\ra$, begin
with some splitting (which must vanish in the thermodynamic limit) and 
then evolve as the inserted flux $\Phi$
changes. They are degenerate and cross at $\Phi=\pi$ since the
Hamiltonian is invariant under time-reversal, but they cannot mix
since the $\hat{x}$-crystal momenta of the two states differ by $\pi$ 
even at this point. 
If the geometry is chosen such that $2\pi \nu L_y = 0
(\mbox{mod }2\pi)$ these two states will no longer exchange places
on threading $\Phi=2\pi$ (see Fig.~\ref{crossing}(b)). 
They no longer cross at $\Phi=\pi$ though they still carry 
relative $\pi$ momentum \cite{footnote2}.

%%%%%%%%%%%%%%%%%%%%%%%%%%%%%%%%%%%%%%%%%%%%%%%%%%%%%%%%%%%%%%%%%%%%%%
%%%%%%%%%%%%%%%%%%%%%%%%%%%%%%%%%%%%%%%%%%%%%%%%%%%%%%%%%%%%%%%%%%%%%%

%The momentum carried by the quasi-degenerate states and the manner
%in which they evolve under flux threading is thus a useful diagnostic
%of translational symmetry breaking. We discuss this further in
%Section \ref{numerics}
%in connection with numerical exact diagonalization studies
%of model Hamiltonians.

\subsection{Local operators to detect broken symmetry states}

In the presence of spontaneous translational symmetry breaking, there
are local operators which can distinguish the different insulating
ground states obtained in the thermodynamic limit by taking linear
combinations of the degenerate momentum eigenstates as $\vert 1 \ra =
(\vert 0\ra+\vert \pi\ra)/\sqrt{2}$ and $\vert 2 \ra = (\vert
0\ra-\vert \pi\ra)/\sqrt{2}$. For example, broken translational
symmetry along say the $\hat{x}$ direction means it can be detected by
some local Hermitian operator $\hat{O}_i$, since $\la 1 \vert
(\hat{O}_i - \hat{O}_{i+\hat{x}}) \vert 1\ra \neq 0$, and similarly
for state $\vert 2\ra$. How does this manifest itself on a finite size
system where such linear combinations $\vert 1\ra, \vert 2\ra$ are not
eigenstates of the Hamiltonian?

To answer this, consider the matrix element $\la 0 \vert \hat{O}_i
\vert \pi \ra$ of the local operator between the eigenstates on the
finite system. Since $\hat{O}_i$ is defined locally, it is not
translationally invariant and such matrix elements will be nonzero in
general on a finite system. However, knowing that $\hT \vert 0\ra =
\vert 0\ra$ and $\hT \vert \pi\ra=-\vert \pi \ra$ we can rewrite this
matrix element in the thermodynamic limit as
\bea
2 (\la 0 \vert \hat{O}_i \vert \pi \ra) &=&
\la 1 \vert \hat{O}_i \vert 1 \ra - \la 2 \vert \hat{O}_i \vert 2 \ra \nn\\
&=& \la 1 \vert (\hat{O}_i - \hT \hat{O}_{i} \hT^{-1}) \vert 1 \ra \nn\\
&=& \la 1 \vert (\hat{O}_i-\hat{O}_{i+\hat{x}}) \vert 1 \ra \neq 0
\eea
Thus, the matrix element of this local operator $\hat{O}_i$ between
states of the quasi-degenerate ground state manifold, $\vert 0\ra$ 
and $\vert \pi\ra$, survives in the
thermodynamic limit and implies translational symmetry breaking. 
The local operator could be, for instance, the
energy (or charge or current) density. 

This is the crucial difference between broken symmetry insulators
and translationally invariant fractionalized insulators dealt
with in the next Section. In the latter case, matrix elements of all local
operators between states forming the quasi-degenerate ground state
manifold vanish in the thermodynamic limit. The system size dependence
of the matrix element of local operators between states forming the
quasi-degenerate ground state manifold thus distinguishes an insulator
with translational symmetry breaking from a uniform fractionalized
insulator. However, constructing such local operators needs some 
knowledge of the kind of broken symmetry, in contrast to our general
conclusions in the earlier Section regarding the momenta and evolution 
of the quasi-degenerate manifold of ground states which does not rely 
on such information. We return to this issue in Section \ref{numerics}.

In this section we have focused on conventional insulating states of a half filled system, and argued that they necessarily break a lattice symmetry. One case however needs to be looked at separately, and that is the case of exactly $\pi$ flux through every 
elementary plaquette, which could be self generated in the thermodynamic limit. Note, this situation can preserve time reversal symmetry, and hence should be admitted in our discussion. It is possible for such a system to be essentially an insulator at
 half filling, although it appears to possess translation symmetry, in that all unit cells appear identical. This issue is resolved  by studying more carefully the meaning of translation invariance --- it turns out that the operators that generate unit
translations do not commute due to the presence of $\pi$ flux in the elementary plaquette. Hence, the smallest mutually commuting translations necessarily enclose an area
equal to two unit cells, and in this sense we obtain unit cell doubling. Note, the 
emergence of $\pi$ flux per plaquette is a property that can be checked with local operators, 
and hence also corresponds to a conventional (non-fractionalized) state.

\section{$Z_2$ fractionalized insulator $\cis$}
\label{Z2insulator}

In the context of the insulating phase of the high temperature
superconductors the question has been raised of whether a Mott
insulator that breaks no symmetries could be obtained at half-filling.
The analogous question for our bosonic system is whether a
translationally invariant insulating state can be realized at
half-filling. Since the hard-core boson state at half-filling may be
viewed as a spin $S=1/2$ system with $S^{\rm total}_z=0$ and $U(1)$
spin rotation invariance, this is
equivalent to asking whether a $S=1/2$ magnet may be in a spin liquid
state. The answer, after several years of work, is {\em yes}, and one
specific route to realizing such an insulator is via $Z_2$
fractionalization.  The properties of such a phase \cite{Senthil_Z2} 
as well as some
microscopic models which realize them are now known.\cite{Microscopicmodels}

The $Z_2$ fractionalized insulator, $\cis$, is a translationally invariant
insulator. It is {\em unconventional} in that supports gapped {\it
fractionally charged} excitations, chargons, which carry electromagnetic
charge $Q/2$ as well as a $Z_2$ Ising charge. These chargons interact 
with a $Z_2$ Ising gauge field in its deconfining phase. The deconfinement 
is reflected in the presence of yet another exotic gapped neutral
excitation, the Ising vortex or ``vison'', which acts as a bundle of
$\pi$ flux as seen by the chargons which carry Ising charge \cite{Senthil_Z2}.

It is known that the $\cis$ phase can be realized at half odd-integer
or integer fillings. {\em Why doesn't the existence of a
translationally invariant $\cis$ insulator at half-integer filling
contradict the earlier theorem for (conventional) insulators?} The
resolution of this apparent paradox is that although these insulators
do not break translational invariance, the ground state of these
fractionalized insulators is not unique in a multiply connected
geometry (in which the flux-threading experiment is carried out).  The
presence of the $Z_2$ vortex, the vison, directly leads to a two-fold
degeneracy of the ground state of the system on a cylinder (four-fold
on a torus). This degeneracy may be viewed as a result of having or
not having a vison threading each hole of the cylinder (torus). Since
the vison is a gapped excitation in the bulk of the system, there is
an infinite barrier for the tunneling of the vison ``string'' out of
the hole of the cylinder (torus) in the thermodynamic limit. Thus the
vison/no-vison states do not mix in the thermodynamic limit which is
crucial to obtaining leading a ``topological degeneracy'' -
i.e. degeneracy which depends on the number of holes in the system.

At this point we introduce the following terminology for the $Z_2$
fractionalized insulators. The translationally symmetric $Z_2$
fractionalized insulator at half-odd-integer filling (integer filling)
will be denoted as $\ciso$ ($\cise$). While in the former case,
translation symmetry of a half filled insulator implies that the state
must be exotic, it is of course possible to have a completely
conventional insulator at integer filling. Nevertheless, a $Z_2$
fractionalized insulator may also exist at integer filling and we
refer to this as ($\cise$). We will see below that these two classes
of exotic insulators are in fact closely related to two classes of
$Z_2$ gauge theories, $Z_2^{odd}$, $Z_2^{even}$ in the
terminology of Ref.~\onlinecite{Moessner02}.

The presence of topologically degenerate states and their evolution
under flux threading, allows us to satisfy the momentum balance
condition. The relevant case to consider is the translationally
symmetric insulator at half filling, on a cylinder with an odd number
of rows. In this case, trivial momentum counting tells us that $2\pi$
flux threading leads to a degenerate state with crystal momentum
$\pi$. We will argue below, this momentum is accounted for in $\ciso$
since flux threading effectively adds a vison into the hole of the
cylinder, which carries crystal momentum $\pi$. 
%For the insulator at
%half-odd integer, which we shall denote $\ciso$ for reasons clarified
%below, this allows for a possible change from one ground state to
%another topologically degenerate ground state which differs in
%momentum by $\pi L_y$, and is thus consistent with the momentum
%counting arguments for $\nu$ being a half-integer. On the other hand,
%for insulators at integer filling, denoted $\cise$, the different
%topologically degenerate ground states carry the same momentum and
%flux threading does not interchange them. The momentum count is thus
%satisfied for fractionalized insulators due to the presence of
%topological degeneracy. While both $\cise$ and $\ciso$ share the same
%excitation spectrum (gapped chargons and visons), they differ in a
%crucial manner under flux-threading which allows for a sharp
%distinction.

\subsection{Effective Hamiltonian for $\cis$}

The effective description of $\cis$ is via a set of gapped
charge-$Q/2$ bosons (chargons) also carrying an Ising charge,
interacting with each other and minimally coupled to an Ising gauge
field in its deconfining phase. In order to place the following
discussion on a more concrete footing we consider a definite
Hamiltonian that can describe such a system, and use it to derive
properties of the states. Since we will be interested in universal
properties that characterize the state, the results themselves are
more general than the particular effective Hamiltonian used. 
The simplest Hamiltonian which can describe a $Z_2$ fractionalized
insulator is: 
\be
H_A(\cis) = H_g + H_m
\ee 
where 
\bea 
H_g\!\!\!&=&\!\!\! - K \sum_{\Box} \prod_{\Box} \sigma^z_{ij} - h \sum_{\la
ij\ra} \sigma^x_{ij} \\ 
H_m\!\!\! &=&\!\!\! - t_b \sum_{\la ij\ra}
\sigma^z_{ij} (b^\dg_i b^\pdg_j e^{- i Q A_{ij}/2} + h.c.) \nn\\
&+& U \sum_i (n_i - 2\bar{N})^2
\label{HZ2}
\eea
where $\sigma^{x,z}$ are Pauli matrices describing the Ising gauge
fields, and $\Box$ denotes the elementary plaquette on a square
lattice. The chargons, created by $b^\dg_i$, are minimally coupled to
the Ising gauge field, as well as to the external vector potential
$A_{ij}$ with electromagnetic charge-$Q/2$.  The second term in $H_m$
describes repulsion between chargons at the same site.

The Hamiltonian (\ref{HZ2}) has a local $Z_2$ 
invariance under
the transformation $b_i \to \alpha_i b_i$ and $\sigma^z_{ij} \to
\alpha_i \sigma^z_{ij} \alpha_j$ where $\alpha_i = \pm 1$. Such gauge
rotations are generated by unitary transformations using the operator
$\hat{G} = \prod_i \hat{G}_i$ with
\be
\hat{G}_i = \exp\big[ i\frac{\pi}{4} (1-\alpha_i) 
(\sum_{j=nn(i)} \sigma^x_{ij} + 2 n_i) \big]
\ee
Local $Z_2$ invariance implies 
that $[\hat{G}_i, H_A(\cis)]=0$.  Since we wish to
work with eigenstates of $H_A(\cis)$ which are invariant under such
gauge transformations, translationally invariant physical states have 
to satisfy 
\be
\hat{G}_i
\vert {\rm phys}\ra = (\pm 1)  \vert {\rm phys}\ra.
\label{phys}
\ee
Let us choose $\hat{G}_i = +1$ everywhere.

It is instructive to first consider the limit $h,U \gg K,t_b$. In this case,
since $h \gg K$, the gauge theory is confining. Depending on the filling, 
it is then possible to show that one
recovers conventional insulating phases such as a uniform band
insulator (for $\bar{N}=$ even integer), or broken 
symmetry states 
such as bond-centered (with $\bar{N}=$ odd integer, and $U \gg h$) or 
site-centered (with $\bar{N}=$odd integer and $h \gg U$) 
charge density wave states. Thus, the above effective
Hamiltonian in this limit is capable of describing well understood 
conventional insulators. 

However, this Hamiltonian has a richer phase 
diagram.
The parameter regime where an exotic fractionalized insulator is expected 
for the above Hamiltonian is easily determined. For $K \gg h$, the Ising
gauge field will be in its deconfining phase, so we can pick \cite{footnote3}
$\sigma^z_{ij} \approx 1$. Similarly, since we are interested in the
insulating phase, let us work in the limit of large chargon repulsion
$U/t \gg 1$ with $2\bar{N}$, which is twice the filling fraction of
the charge $Q$ bosons, being an integer. In this limit, it is clear
that it is energetically favorable to also set the
chargon number $n_i=2\bar{N}$ at each site (which
is possible since $2\bar{N}$ is an integer) as a starting point to
understand the insulator.  The density of charge-$Q$ bosons in the
insulator is just $\nu= \bar{N}$, and $\nu$ could thus either be an
integer or a half-odd integer in the $\cis$ phase corresponding to
even/odd integer values of $2\bar{N}$.

In the above regime of parameters, the system clearly has a charge gap
${\cal O}(U)$ for adding an $n_i$ particle (chargon) which is a
charge-$Q/2$ and Ising-charged excitation that can propagate freely
(since the gauge field is deconfined). It also has an energy gap
${\cal O}(K)$ to changing $\sigma^z_{ij} \to -1$ on a bond which
changes $\prod_\Box \sigma^z_{ij} \to (-1)$ on adjacent plaquettes
corresponding to creating gapped visons. It thus describes an exotic
insulator.

\subsection{Flux threading in $\cis$}
Below, we will consider the effect of threading $2\pi$ flux on the
$Z_2$ fractionalized insulators in the cylindrical geometry, using the
effective Hamiltonian (\ref{HZ2}). This will be done in two steps. We
first consider the limit of being deep in the fractionalized phase
(i.e. set the vison hopping to zero; $h=0$ in equation
\ref{HZ2}) where it can be easily argued that $2\pi$ flux threading
leads to the insertion of a vison through the hole of the
cylinder. The momentum balance argument then allows us to read off the 
crytal momenta of the visons in the different situations. Then, we
turn back on a finite vison hopping $h\neq 0$, and use continuity
arguments to conclude that these crytal momenta assignments remain unchanged.

%From the trivial momentum counting, we know that threading flux
%$\Phi_0=2\pi$ in the system changes the crystal momentum by $2\pi\nu
%L_y$. How can we account for this momentum change in the effective
%theory for $\cis$? 

\noindent{\bf Flux threading with static visons} 
Consider at first the limit of being deep in the fractionalized phase
$h/K \to 0$ by setting $h=0$ identically (i.e. no vison hopping), so
that we can choose $\sigma^z_{ij}=1$ everywhere \cite{footnote3}.  Let
us adiabatically thread flux $2\pi$ in the $\hat{y}$ direction for the
above system on a cylinder such that the starting from the initial
eigenstate $\vert \Psi(0)\ra$ in the absence of flux, the final state
\bea
\vert \Psi(T)\ra &=& \cU_T \vert \Psi(0)\ra \\
\cU_T&=&{\cal T}_t \exp\left(-i \int_0^T H_A(\cis,t) dt\right)
\eea
where ${\cal T}_t$ is the
time-ordering operator. 
We can go to the $A_{ij}=0$ gauge by making a
unitary transformation $H_A(\cis,T) \to \cU_G H_A(\cis,T) \cU_G^{-1}
\equiv H_0(\cis)$ (corresponding to zero flux).

\begin{figure}
\begin{center}
\vskip2mm
\hspace*{0mm}
\centerline{\fig{3.0in}{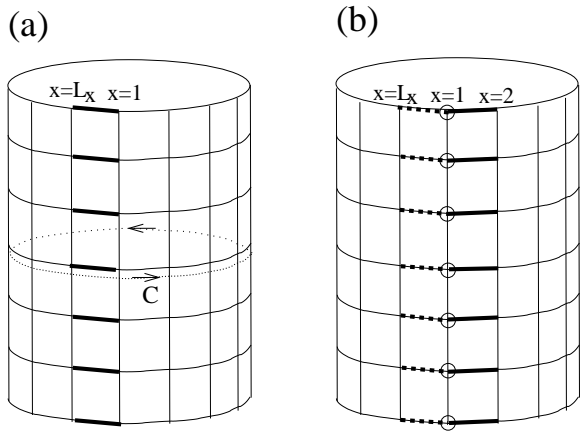}}
\vskip2mm
\caption{(a) Schematic figure showing a vison threading the hole
of a cylinder in the absence of vison tunneling terms. The dark
(light) bonds correspond to $\sigma^z_{ij}=-1$ ($\sigma^z_{ij}=+1$).
We can detect the presence of the vison by
evaluating the Wilson loop operator $\Pi_C \sigma^z_{ij}$ along
the contour $C$ taken around the cylinder. (b) The translation
operator along the $\hat{x}$ direction moves the dark ($\sigma^z_{ij}=-1$)
bonds by one lattice spacing from $(L_x,1)$ to $(1,2)$ at each $y$.
This is accomplished equivalently by acting with the gauge transformation
operator $G_i$ (which changes the sign of $\sigma^z_{ij}$ on all bonds
emanating from $i$) acting on each of the circled sites.}
\label{cut}
\end{center}
\end{figure}

Since the chargons carry a charge $Q/2$, the $hc/Q$ flux quantum
threading the cylinder appears as an Aharonov-Bohm flux of $\pi$
for the chargons. The gauge transformation which returns the
Hamiltonian to its original form
thus also acts on the Ising gauge fields to remove 
this extra $\pi$-flux. Hence we have,
\bea
\cU_G&=&\cU_\phi \cU_\sigma, ~~{\rm with} \\
\cU_\phi&=&\exp(i \frac{\pi}{L_x} \sum_i x_i \hat{n}_i) \\
\cU_\sigma&=&\prod_{ij\in{\rm cut}} \sigma^x_{ij}
\eea
and 'cut' refers to the set of links for which $x_i=L_x,x_j=1$ (shown
in Fig.~\ref{cut}). Thus, the final state in the $A_{ij}=0$ gauge is $\vert
\Psi_f\ra = \cU_\phi \cU_\sigma \cU_T \vert \Psi_i \ra$.
%%%%%%%%%%%%%%%%%%%%%%%%%%%%%%%%%%%%%%%%%%%%%%%%%%%%%%%%%%%%%%%%%%%%%%
%%%%%%%%%%%%%%%%%%%%%%%%%%%%%%%%%%%%%%%%%%%%%%%%%%%%%%%%%%%%%%%%%%%%%%
Since the
system is an insulator, the final state on threading flux $\Phi_0$
must be one of the states which forms part of the degenerate ground
state manifold in the thermodynamic limit.

Let us define the loop operator $W_C=\prod_{C} \sigma^z_{ij}$ where
the loop $C$ is taken around the cylinder (see Fig.~\ref{cut}). Clearly, since
$h=0$, this operator commutes with the Hamiltonian (\ref{HZ2}). 
We can use this
operator to check whether there is a vison through the hole of the
cylinder. Namely, if we are in a (reference) state with
$\sigma^z_{ij}=1$ everywhere, then $W_C=1$ and this is the no-vison
state $\vert v=0\ra$. If on the other hand $W_C = -1$ for each loop $C$ around
the cylinder, we conclude that there must be a vison threading the
hole, and we shall refer to this as $\vert v=1\ra$. Let us evaluate
$W_C$ for the two states $\vert \Psi_{i,f}\ra$ above. We easily find
$W^i_C=1$ in the initial state.  In order to find the eigenvalue of
$W_C$ in the final state, we note that since $h=0$, the initial
assignment of 
$\sigma^z_{ij}$ does not time evolve, and we only need to evaluate the
effect of the unitary transformation $\cU_G$ on $W_C$. This yields the
result that the eigenvalue in the final state is:$W^f_C=-1$. 
%due to acting
%with the operator $\cU_\sigma$ on the time evolved initial state.

Thus, for $h=0$, threading a $2\pi$ flux adds a vison to the hole of
the cylinder and interchanges the two ground states on the cylinder,
$\vert v=0\ra \leftrightarrow \vert v=1\ra$. (In fact, in the absence
of dynamical matter fields, i.e. $t_b=0$, the operator $U_\sigma
\equiv \prod_{ij\in{\rm
cut}} \sigma^x_{ij}$ commutes with the Hamiltonian and can be viewed
as the ``vison creation operator'' introducing a vison into the hole
of the cylinder and changing the sign of $W_C$.)  

Momentum balance then tells us that adding a vison into the hole of
the cylinder must change the momentum of the system by $2\pi\nu
L_y$. The only situation where this is a nontrivial crystal momentum
is for the case of $\ciso$ on a cylinder with an odd number of rows
($L_y$ odd). Then we expect the two states $\vert v=0\ra$ and $\vert
v=1\ra$ to differ by crystal momentum $\pi$. In all other cases,
i.e. for $L_y$ even, or of $\cise$, the vison carries no momentum. 

As we shall see in the next subsection, this is
consistent with a direct computation of the vison momentum in the pure
Ising gauge theory. 
We now switch back on the vison hopping $h \neq 0$
and ask how these conclusions might be affected.

% (ignoring the dynamics of gapped chargons), and the
% two insulators, with $2\bar{N}$ being an even/odd integer will be
% referred to as $\cise/\ciso$ respectively.

\noindent{\bf Vison state in the presence of dynamical gauge
fields:}
%Next, starting from the case with $h=0$, let us ask what happens if
%we

Turning on a nonzero $h$, gives dynamics to the gauge field. In this
case the loop product $W_C$ no longer commutes with the Hamiltonian,
we cannot use its eigenvalues to label the states. Let us first see
what effect this has on $\cise$. The two states $\vert v=0\ra$ and
$\vert v=1\ra$ both carry zero crystal momentum, and will now mix to
give eigenstates of the Hamiltonian. Thus, on flux threading there is
no level crossing - threading a $\Phi_0$ flux returns us to the
original ground state. For $\ciso$ on a cylinder with even $L_y$, the
two low lying states carry zero crystal momentum, and a similar
conclusion applies. 

The situation is more interesting for $\ciso$ on a cylinder with an
{\em odd} $L_y$. Now, the states $\vert v=0\ra$ and $\vert v=1\ra$
cannot mix since they carry different momenta. Thus, even in the
presence of $h\neq 0$ (so long as we remain in the same phase), we can
continue to distinguish them and we can continue to label them as
no-vison/vison states by their momentum, although they are not
eigenstates of the $W_C$ operator any longer. In this case, the
crossing of the two states on threading a $\Phi_0$ flux continues to
occur, since the crystal momentum must change by $\pi$ in order to
satisfy momentum balance. Thus we may conclude that for the case of
$\ciso$ on an odd length cylinder, the two degenerate ground states
(no-vison and vison through the hole of the cylinder) differ by
crystal momentum $\pi$. This is the result of the momentum balance
argument applied to $Z_2$ fractionalized insulators.
We will sometimes simply refer to this result for $\ciso$
as ``the vison
carrying momentum $\pi$ per row of the cylinder'',
omitting to point out each time that the vison in question lives
in the hole of the cylinder.

The above result is consistent with the vison
momentum computed using: (i) the pure Ising gauge theory 
(as shown in the next subsection), (ii) variational 
wavefunctions for $Z_2$ spin liquids (as shown in
Section \ref{Z2insulator}~D), and, (iii)
arguments presented for short-range dimer models \cite{Bonesteel89}
for $\ciso$.

A side result of this analysis of identifying the vison crystal
momenta in various situations is an unambiguous way of distinguishing
fractionalized states from states with translation symmetry breaking
for the half filled insulator. This is described in detail in Section
\ref{numerics}.

%The two insulators for $\nu$ integer ($\cise$) or half odd-integer
%($\ciso$) can thus be distinguished depending on the way in which
%levels cross on flux threading. As discussed in Section XXX, this can
%be a useful way to numerically check for $Z_2$ fractionalization, and
%distinguish it from translational symmetry breaking insulators.

\subsection{Vison momentum computed directly in the pure Ising gauge theory}

In order to check our deduction about the vison
momentum, let us directly compute this quantity in a pure $Z_2$ Ising
gauge theory without dynamical matter fields. If
the charge gap in the insulator is large, this is the effective
description of the insulator $\cis$. Namely, in the limit $U \gg t_b$
in the Hamiltonian (\ref{HZ2}) and for integer values of
$2\bar{N}$, a good caricature of the insulating state is to set
$n_i=\bar{N}$ at each site and only consider fluctuations of the Ising
gauge fields. This reduces the constraint on the physical Hilbert
space to

\bea
\hat{G}^{red}_i \vert {\rm phys} \ra &=& (-1)^{2\bar N} \exp\big[
i\frac{\pi}{4} (1-\alpha_i) \sum_{j=nn(i)} \sigma^x_{ij}\big] \vert
{\rm phys} \ra \nn \\
&=& \vert {\rm phys}\ra
\eea
or equivalently, focussing only on the non-trivial case of
$\alpha_i=-1$,
\be
 \prod_{j=nn(i)}
\sigma^x_{ij}= (-1)^{2 \bar{N}}
\label{puregauge}
\ee
in the subspace of physical states.

Again, if we begin with $h=0$, one ground state $\vert v=0\ra$ of the
gauge theory on a cylinder may be obtained as the reference state
$\sigma^z_{i}=1$ projected into the physical subspace, and for this
one has the loop operator $W_C=1$. A second (degenerate) state may be
obtained by acting on this ground state with 
\be
V_{L_x,1}^\dg =
\prod_{ij\in{\rm cut}} \sigma^x_{ij}
\label{visoncreation}
\ee
which commutes with the $H_g$
for $h=0$. The subscripts $(L_x,1)$ on $V^\dagger$ are a
mnemonic for the
column on which the $\sigma^x$ operators act as shown in Fig.~\ref{cut}(a).
The resulting state has $W_C=-1$. Let us compute the
momentum of these two states. Clearly, the state $\vert v=0\ra$ has
zero momentum since it is translationally invariant by
construction. To compute the momentum of the second state,
we first note (see Fig.~\ref{cut}(b)) that 
\bea
\hT V^\dg_{L_x,1} \hT^{-1} &=&  V^\dg_{1,2} \\
&=&  \left[ \prod_{i} \big(\prod_{j\in nn(i)}
\sigma^x_{i,i+\hat{x}} \big) \right]
V^\dg_{L_x,1}
\eea
where the sites $i$ have $x_i=L_x$ and 
correspond to the circled sites in Fig.~\ref{cut}(b). Using
the constraint in Eq.~\ref{puregauge}, this reduces to
\be
\hT V^\dg_{L_x,1} \hT^{-1} = 
= \left( \prod_{i} (-1)^{2\bar{N}}
\right) V^\dg_{L_x,1} = \exp(i 2\pi \bar{N} L_y) V^\dg_{L_x,1}.
\ee
Thus, for the second state, $\vert v=1\ra = V^\dg_{L_x,1} \vert
v=0\ra$, acting with the translation operator leads to 
\bea
\hT \vert v=1\ra &=& 
\left(\hT V^\dg_{L_x,1} \hT^{-1}\right) \hT \vert v=0\ra \\
&=& \exp(i 2\pi \bar{N} L_y) V^\dg_{1,2} \vert v=0\ra \\
&=& \exp(i 2\pi \bar{N} L_y) \vert v=1\ra
\eea
%From the constraint that we restrict
%to the physical Hilbert space, we know $G^{red}_i \vert v=1\ra = \vert
%v=1\ra$ for every site $i$. 
%Since the product
%$\prod_{i} (-1)^{2 \bar{N}} = (-1)^{2 \bar{N} L_y}$, we have 
%\be
%\hT \vert
%v=1\ra = \exp(i \pi 2\bar{N} L_y) \vert v=1\ra
%\ee
In other words, 
(i) for even $\bar{2 N}$, namely in $\cise$, the state $\vert v=1\ra$
carries zero crystal momentum and , (ii) for odd $\bar{2 N}$, namely in
$\ciso$, the vison state $\vert v=1\ra$ carries momentum $\pi L_y$.

As before, we can now turn on a nonzero field $h$. In $\cise$, the two
ground states will mix and split in a finite system, since they carry
the same momentum quantum number. The same is true for $\ciso$ with
even $L_y$. However, for $\ciso$ with odd $L_y$, the two ground states
carry relative momentum $\pi$, thus they cannot mix even on a finite
system with nonzero $h$ and can be distinguished by their
momentum. 

Finally, we may introduce dynamical matter fields. Although the
operator in Eqn. \ref{visoncreation} no longer can be identified as a
vison creation operator, the low energy structure of the system,
i.e. topological degeneracies will not change as long as we are in the
same phase. Also, since the crystal momenta of these low lying states
on the cylinder can only be one of $0,\pi$ (from time reversal
invariance), continuity requires that the crystal momentum assignments
made before for the low lying states continue to hold in the presence
of dynamical matter fields. This constitutes a direct check of the
results deduced above using momentum balance arguments.

\subsection{Momentum computation from variational wavefunctions for the 
vison}

So far, we have discussed bosonic models for the insulators
$\cis$. However, as mentioned earlier, we can view a hard-core boson
as a $S=1/2$ spin, and the insulating state $\ciso$ with $\bar{2 N}$
an odd integer as a $Z_2$ fractionalized spin liquid insulator.  Such
spin liquid insulators have long been of interest in connection with
frustrated magnets and the high temperature superconductors. The
$Q=1/2$ chargon excitations in the bosonic language correspond to
$S=1/2$ excitations (called spinons) in the spin liquid. What do the
visons in $\ciso$ correspond to?

To answer this, we note, following Anderson\cite{Anderson87}, 
that one can represent
of the ground state wavefunction for spin liquids with short-range
antiferromagnetic correlations by ``Gutzwiller projecting'' a
superconducting wavefunction, i.e. restricting to configurations
with a fixed number of electrons per site. Such a picture also
emerges from mean-field studies of frustrated magnets using a
fermionic representation for the spins. This suggests that perhaps
excitations of the spin liquid may also be related to excitations
in the superconductor. Following this line of thought, the $S=1/2$ 
spinon in the spin liquid may be viewed as a projected
Bogoliubov quasiparticle of the superconductor. Similarly it is natural 
to expect that the $hc/2e$ vortex in the superconductor becomes 
the vison \cite{wavefunctions}.

We can check this possibility by computing the momentum of a projected
$hc/2e$ BCS vortex threading the cylinder, with odd/even number of
electrons at each site, and seeing if it agrees
with the results for $\ciso/\cise$ obtained above. To do this, we write the
BCS state (with total electron number $N_e$) as
\be
\vert BCS(N_e) \ra = \left(\sum_\bk \phi_\bk c^\dg_{\bk\uparrow} 
c^\dg_{-\bk,\downarrow} 
\right)^{N_e/2},
\ee
where $\phi_\bk$ denotes the internal pair state of the Cooper pair
formed by $(\bk,\uparrow)$ and $(-\bk,\downarrow)$.  which carries
zero center of mass momentum. 

To get the spin liquid state at half-filling, we have to
choose $N_e=N$, the number of lattice sites, and Gutzwiller project
this state by acting with the operator $P_G = \prod_i (1-n_{i\uparrow}
n_{i\downarrow})$ which eliminates configurations in which two
electrons occupy the same site. The variational ansatz for the spin
liquid ground state is then $P_G \vert BCS(N) \ra$, and it is a
translationally invariant state with zero momentum. To construct a BCS
$hc/2e$ vortex threading the cylinder, we need to pair states with
$(\bk+\bq/2,\uparrow)$ and $(-\bk+\bq/2,\downarrow)$ with
$\bq=(2\pi/L_x) \hat{x}$ and amplitude $\phi_\bk$, this leads to the
$N$-particle vortex state $\vert hc/2e (N)\ra$ carrying a momentum of
$\bq$ per pair, or a total momentum $(2\pi/L_x) (N/2)$. Setting $N=L_x
L_y$, we see that the momentum of this state is just $\pi L_y$. The
projection operator $P_G$ commutes with the translation operator. Thus
the trial vison state $\vert v\ra=P_G \vert hc/2e (N)\ra$ has a
momentum $\pi L_y$ in agreement with earlier arguments for the $\ciso$
insulator. 

With an even number of electrons at each site, the vison wavefunction
carries no crystal momentum. This is consistent with our earlier result 
for $\cise$.

\section{Identifying $Z_2$ Fractionalized States in Numerical
Experiments.}
\label{numerics}

Numerical investigation of microscopic models, for example exact
diagonalization studies \cite{Misguich_numerics,Sheng04}, 
are an important tool in finding new states
of matter such as states with $Z_2$ topological order. In this context
it is important that reliable diagnostics be available for the
identification of these states in the system sizes that can currently
be solved on the computer. This question may seem straightforward in
principle, the four fold topological degeneracy of the $Z_2$ states on
a torus that are indistinguishable by any local operator seem to
provide a unique prescription. However, in practice there are several
potential problems. First, since the numerical simulations are
performed on finite sized systems, states that are degenerate in the
thermodynamic limit are only approximately so in these systems. The
problem is particularly severe when gapless $Z_2$ charged matter fields are
present, in which case the splitting between topological sectors that
differ by the presence of a vison can be large and go down to zero
only algebraically with system size (in contrast to the exponentially
small splitting in the absence of such gapless gauge charged
matter fields). Secondly, if
degeneracies arise as a result of broken translation symmetry, rather
than topological order, the relevant order parameter for this
translation symmetry breaking may be hard to identify, and hence we
would like to have available a prescription for distinguishing such
states even if the order parameter is not known. Below we use insights
from the momentum balance arguments to resolve both of these
issues. Indeed we will see that the analysis of the previous sections,
with their focus on finite sized systems and crystal momentum quantum
numbers, are ideally suited to addressing these questions. We begin by
addressing the second of these two questions first - i.e. given a set
of states comprising the low energy manifold of the system as the
thermodynamic limit is approached, how does one distinguish
topologically ordered states from a conventional translation symmetry
broken state?

\subsection{$Z_2$ Topological Order vs. Translation Symmetry Breaking}
Consider a system of bosons on a lattice at half filling (or
equivalently a spin 1/2 system with one spin per unit cell). As
discussed previously, a translationally invariant insulating phase
implies the presence of topological order (although it is possible to
have topologically ordered phases that also break translation
symmetry). Assume that a group of low lying states have been
identified --- under what conditions can we associate these with the
degeneracies assocated the $Z_2$ topological order, rather than with
low lying states leading to translation symmetry breaking?

First consider the system in the cylindrical geometry with an odd
number of rows ($L_y$ odd, $L_x$ even in Fig.~\ref{cylinder}). Then,
flux threading ensures that we will have two low lying states with
crystal momentum $P_x=0$, $P_x=\pi$ which are interchanged on
threading $2\pi$ flux. This is irrespective of whether the system
is heading towards translation symmetry breaking or towards a $Z_2$
topologically ordered state in the thermodynamic limit. Therefore this
setup is not particularly useful for discussing for distinguishing the
two states. One may also consider the toroidal geometry for an even
$\times$ odd system, however, this can potentially frustrate certain
patterns of
translation symmetry breaking in a half filled system and hence we do
not consider it further here.

Now consider the system in a toroidal geometry, but with both $L_y$,
$L_x$ {\it even}. Now, we have shown earlier that a $Z_2$
topologically ordered state will have four low lying excitations, all
with zero crystal momentum in this geometry. A conventional
translation symmetry breaking state on the other hand will invariably
have at least one state in the low energy manifold that carries
nonzero crystal momentum, in addition to a zero crystal momentum
state. This follows almost by definition, in order to build up a
translation symmetry breaking state one needs to make a linear
combination of states with different crystal momenta. This then is a
precise way to tell apart a $Z_2$ topological state from a more
conventional translation symmetry breaking state, which just requires
a correct identification of the low energy manifold and the crystal
momenta of these states. The translationally symmetric topologically
ordered state is present if there are four low lying states with zero
crystal momenta. If topological order coexists with translation
symmetry breaking, then too this quadruplet of zero momentum states
persist, although other quadrupled states with different crystal
momenta will be present in the low energy manifold. This is true
for both $\cise$ and $\ciso$.

Recent exact diagonalization studies studies of a multiple spin exchange model on a triangular lattice
have found signatures of an interesting new spin state, which has been proposed to be a topologically ordered spin liquid phase in a certain regime of parameters 
\cite{Misguich_numerics}. Let us apply the method of distinguishing topological order 
from broken translational symmetry discussed above, to these states.

In the parameter regime of interest, the system in Ref.~\onlinecite{Misguich_numerics} 
was argued to be heading, in the thermodynamic limit, towards a spin gapped phase 
without long-range magnetic order. 
Furthermore, a set of three spin singlet states which appear to become degenerate
with the ground state with increasing system size were identified.  
The authors were unable to find simple valence bond (eg. nearest neighbour) crystal states that would  lead to degenerate ground states with the quantum numbers
(crystal momenta, rotations, reflections) of these low lying states. Hence they identified this 
apparent four-fold degeneracy with the degeneracy arising from  topological
order of a $Z_2$ fractionalized spin liquid, such as described by Eqn \ref{HZ2}, on a torus.  While this would be a very interesting result, we can ask if the quantum numbers of these nearly degenerate states are consistent with those of a vison in an 
odd Ising gauge theory, that we have 
derived earlier.  
However, the three excited states which appear to become degenerate
with the ground state carry {\it nonzero} crystal momentum on an 
$6\times 6$ lattice \cite{Misguich_numerics}. This is in disagreement with
our conclusion regarding vison states in a $Z_2$ fractionalized phase \cite{footnote.mse},
namely, that
they carry zero crystal momentum on even$\times$even lattices.
We therefore conclude that this interesting identification of $Z_2$ topological order
in these systems does not stand up to detailed scrutiny. The actual nature of the phase being approached by these systems then remains an open question, especially since an extensive search of conventional broken symmetry states in Ref.\onlinecite{Misguich_numerics} did 
not yield a candidate phase. The remaining possibilities are perhaps a conventional translational 
symmetry broken valence bond crystal phase, 
involving non-nearest  neighbour dimers, some other more exotic fractionalized state, 
or that the all the low lying states associated with the broken symmetry have 
not been identified as a consequence of finite size effects.
Note, the 
evolution of these states under flux
threading which they have studied on odd$\times$even and 
even$\times$even lattices is also consistent with
a conventional broken symmetry state. 

\subsection{Eliminating the Spinon Contribution to Vison Splitting}
In this subsection we will utilize the flux threading proceedure to
find a way of eliminating the splitting between the topologically
`degenerate' states that arises from the presence of $Z_2$ charged
matter fields. While in the thermodynamic limit, a $Z_2$
fractionalized system must posess a four fold ground state degeneracy
on the torus and a two fold ground state degeneracy on the cylinder,
in a finite system this spiltting may be so large that it makes the
identification of the low energy manifold problematic. In this
subsection we will utilize the flux threading proceedure to find a way
of eliminating the part of the splitting between these states that
arises from the presence of $Z_2$ charged matter fields. 

In order to study the problem in more detail consider a finite sized
system in the cylindrical geometry that is heading towards a $Z_2$
fractionalized insulating state. Consider first the situation on an
even$\times$even lattice. The low energy manifold consists of a pair
of states that eventually become degenerate in the thermodynamic
limit, but at this stage have a finite splitting $\Delta E$. The
spitting arises from two sources: first there is vison tunneling, that
mixes the zero and one vison states, which acting alone, would lead to
a splitting of $\Delta E_{\rm hop}$. Since this involves a gapped
vison (with gap $\epsilon$) hopping across the entire height $L_y$ of
the cylinder, one would expect this to be exponentially small in their
product i.e. $\Delta E_{\rm hop} \propto \exp(-cL_y\epsilon)$, where
$c$ is a constant. The second contribution to the splitting arises
from the presence of matter fields that carry gauge charge. Clearly,
the presence or absence of a vison will affect the propagation of these
particles and in the absence of vison tunneling will give rise to an
energy splitting $\Delta E_{\rm mat}$. With gapped $Z_2$ gauge charged
matter fields (with a gap $e$), clearly this splitting will require
virtual processes where the gapped particle goes once around the
cylinder which implies $\Delta E_{\rm mat} \propto \exp(-c'L_x
e)$. The total splitting is easily seen to be: $\Delta E =
\sqrt{\Delta E_{\rm hop}^2+\Delta E_{\rm mat}^2}$. Thus, in situations
like the one described above, where both visons and gauge charged
matter have a healthy gap, finite sized system studies can in practice
isolate the low energy multiplet that leads to topological degeneracy
in the thermodynamic limit.\cite{Senthil_Z2} 
Note that the presence of
gapless matter fields which are {\em gauge neutral} do not affect these
conclusions. Further, atleast in principle, the topological degeneracies
described here, with splittings which are exponentially small
in system size can be separated from low-lying modes of the gauge
neutral excitations whose splitting scales inversely with system
size.

However, if the {\em gauge charged} matter is gapless (e.g. if there are
fermionic $Z_2$ gauge charged excitations with a Dirac spectrum that
often appear in mean field theories of spin liquids) then the
splitting of the zero and one vison states are no longer exponential
in the perimeter size of the system, but only a power law i.e. $\Delta
E_{\rm mat} \propto L_x^{-\eta}$, where $\eta>0$, and this dominates
$\Delta E$. This is potentially
a serious problem since in a finite sized system the splitting is very
likely to be large and and also hard to distinguish from the low
energy states arising from the gapless fermions, which also have
energies that vanish as the inverse size of the system. Below we will
prove that in the presence of $\pi$ flux (i.e. antiperiodic boundary
conditions for the unfractionalized bosons/magnons) the matter
contribution to the vison splitting is switched off! Essentially this
arises because the $Z_2$ charged particles, which also carry half a
unit of charge, see the antiperiodic boundary conditions as a flux of
(say) $-\pi/2$. Adding a vison then implies a flux of
$\pi/2$. However, since these two situations are related by time
reversal symmetry, the energy contribution from the matter fields in
these two cases is identical - which implies that the splitting arises
solely due to vison tunneling, which can be made exponentially small.

In order to show that at a flux of $\pi$ the matter contribution to
the vison splitting vanishes $\Delta E_{\rm mat}=0$, we adopt the
following procedure. We consider the system (heading towards a $Z_2$
fractionalized state) on an even$\times$even cylinder, and consider
first the limit where the vison hopping is turned off. Then, the splitting
of the low energy states occurs entirely because of the gauge charged
matter contribution. We now consider introducing a vison through the
hole of the cylinder and argue below that exactly at flux $\pi$ the two
states are exactly degenerate (even in a finite system). Since the vison
hopping has been tuned to zero, the remaining source of splitting
(arising from the gauge charged matter fields) must also be zero at
this point. We can then reintroduce the vison hopping, and the states
at flux $\pi$ will now be split, but the splitting occurs entirely
from vison tunneling.

Let us now show that in the absence of vison tunneling, an exact
degeneracy occurs at $\pi$ flux. We consider for definiteness the
model in Eqn.(\ref{HZ2}) - although it contains gapped matter fields
with $Z_2$ gauge charge, the conclusions simply show that all matter
field contributions cancel at this special flux, and hence can be
easily extended to the case of gapless matter fields as well. We
consider the limit of vanishing vison tunneling (i.e. $h=0$ in
Eqn. \ref{HZ2}). We start with zero flux through the cylinder and
consider inserting, adiabatically, a flux of $2\pi$.

Then, as argued in Subsection
\ref{Z2insulator}, inserting a $2\pi$ flux leads to
insertion of a vison. Now, at zero flux, the two low lying states can
be classified in terms of vison number, since the vison hopping has
been set to zero. The vison number is measured by the operator
$W_C=\prod_{C} \sigma^z_{ij}$ where the loop $C$ is taken around the
cylinder (see Fig.2). Clearly, since $h=0$, this operator commutes
with the Hamiltonian, and the two low lying states can be labelled
with the eigenvalues of $(1-W_C)/2$, i.e., the vison number. 
The splitting between these levels arises
entirely from the gauge charged matter fields. On flux threading,
these two states must then interchange - since threading $2\pi$ flux
inserts a vison in this limit. This means that the two levels have to
cross at some point (or more generally at an odd number of points) as a
function of flux. Now, time reversal symmetry tells us that if there
is a crossing point at flux $\phi$, then there must be one also at the
point $2\pi - \phi$. Thus, in order to arrange for an odd number of
crossing to ensure the levels do interchange, we need that there is
always a crossing at flux $\pi$. Thus, the two states with vison and
no vison are exactly degenerate at this value of the flux and hence we
conclude that the splitting from the gauge charged matter fields
vanishes at this value of flux. Now, turning on the vison hopping
$h\neq 0$ will lead to a finite splitting even at $\pi$ flux, but this
splitting arises entirely from vison tunneling and hence at this value
of the flux we have $\Delta E =\Delta E_{\rm hop}$ and hence vanishes
exponentially in the width of the system. These arguments can easily
be taken over to the toroidal geometry as well.

We note that this result is useful even in the study of SU(2)
symmetric spin liquid states \cite{Misguich_numerics}, where although the
intoducing the $\pi$ flux will require breaking the SU(2) symmetry,
this only occurs along one row of the cylinder (e.g. changing the sign
of the exchange constants for the $S_x S_x$ and $S_y S_y$
interactions), and hence may be viewed as a fairly weak perturbation
away from full SU(2). It should also be useful in projected
wavefuntion studies, especially in establishing the existence of $Z_2$
fractionalized states with excitations that have a Dirac dispersion
\cite{wavefunctions}. 

Finally we note that while turning on a flux of $\pi$ is effective in
cancelling the splitting arising from dynamical matter fields, the
splitting from vison tunneling can be cancelled in a like manner by
considering a cylinder with an odd number of rows at half filling,
where the vison and no vison states differ by crystal momentum $\pi$
and hence do not mix. When both these processes are active we expect
the vison and no-vison states to be exactly degenerate.Indeed, this
is borne out by the observation that
for a cylinder with an odd number of rows at half filling, when the
threaded flux reaches $\pi$ there is always a level crossing just from
momentum balance arguments and time reversal symmetry (see Fig.~2(a)).

\section{Momentum Balance for Fermi Liquids}
\label{conventionalFL}
\subsection{Conventional Fermi Liquids}
Let us first briefly review the momentum balance argument due to
Oshikawa \cite{Oshikawa00} for conventional Fermi liquids where it leads
to Luttinger's theorem. Consider fermions with charge $Q$ and spin
$\uparrow, \downarrow$ at a filling per site of $\nu_\uparrow =
\nu_\downarrow=\nu$. Now consider flux threading in the cylindrical
geometry of Fig.~\ref{cylinder} with $L_x$ columns and $L_y$
rows. We imagine threading unit flux $\Phi_0=hc/Q$that only couples to
the $\uparrow$ spin fermions. Via trivial momentum
counting this proceedure can be seen to impart a crystal momentum of:
\be
\Delta P_x = 2\pi \nu_\uparrow L_y
\ee
Similarly, one could imagine performing the flux threading with the
cylinder wrapped along  the perpendicular direction which would yield
a crystal momentum change:
\be
\Delta P_y = 2\pi \nu_\uparrow L_x
\ee

Now in the regular Fermi liquid phase, this crystal momentum imparted
during flux threading is accounted for entirely by quasiparticle
excitations that are generated near the Fermi surface. Using the fact that
long lived quasiparticles exist near the Fermi surface, and the fact
that the Fermi liquid is adiabatically connected to the free Fermi gas,
the quasiparticle population $\delta n_p$ excited during the flux
threading proceedure can be worked out. Clearly, flux threading for
noninteracting fermions will lead to a uniform shift of the Fermi sea
by $\Delta p_x = 2\pi/L_x$ from which the quasiparticle distribution
function can be determined. Indeed all of these excitations are close
to the Fermi surface, which is required in order to apply Fermi liquid
theory. The total crystal momentum carried by these excitations can be
written as:
\be
\vec{\Delta P} = \sum_p \delta n_p \vec{p}
\ee
It is convenient to first evaluate this expression neglecting the
dicrete nature of allowed momentum states in a finite volume system
and treating the shift in the Fermi sea $\delta p_x =2\pi/L_x$ as
infinitesimal. This yields:
\be
\Delta P_x = \oint_{FS} p_x \frac{\vec{\delta p}\cdot \vec{dS_p}}{\frac{2\pi}{L_x}\frac{2\pi}{L_y}}
\ee
where $\vec{dS_p}$ is a vector normal to the Fermi Surface, and the
integral is taken around the Fermi surface. Using Gauss divergence
theorem this can be converted into an integral over the Fermi volume
which yields:
\be
\Delta P_x = \delta p_x \int_{FV} \frac{dV}{\frac{2\pi}{L_x}\frac{2\pi}{L_y}}
\ee
thus, 
$$
\Delta P_x =
\frac{2\pi}{L_x}\frac{V^{\uparrow}_{FS}}{\frac{2\pi}{L_x}\frac{2\pi}{L_y}}
$$
A more careful derivation that keeps track of the discreteness of the
allowed momenta gives the same result. Clearly the relevant Fermi
volume that enters here is that of the up spins. Below we assume for
simplicity that both the $\uparrow$ and $\downarrow$ spins are at
equal filling and so $\nu_\uparrow = \nu_\downarrow =\nu$ and
$V_{FS}^{\uparrow}=V_{FS}^{\downarrow}=V_{FS}$. Equating the results
from the trivial momentum counting, and the momentum counting above
for the Fermi liquid (up to a reciprocal lattice vector) yields:
\begin{eqnarray*}
2\pi\nu L_y &=& \frac{2\pi}{L_x}\frac{V_{FS}}{(2\pi)^2}L_x L_y + 2\pi m_x\\
2\pi\nu L_x &=& \frac{2\pi}{L_y}\frac{V_{FS}}{(2\pi)^2}L_x L_y + 2\pi m_y
\end{eqnarray*}
where $m_x$ and $m_y$ are integers and the two equations above are
obtained from threading flux in the $x$ and $y$ directions. These
equations can be rewritten as:
\begin{eqnarray}
N - L_x L_y\frac{V_{FS}}{(2\pi)^2}&=&  L_x m_x\\
N - L_x L_y\frac{V_{FS}}{(2\pi)^2}&=&  L_y m_y
\label{balancefl}
\end{eqnarray}
where we have introduced the particle number $N=L_x L_y \nu$, an
integer. In order to obtain the strongest constraint from these
equations we consider a system with $L_x$, $L_y$ mutually prime
integers (no common factor apart from unity). Then, $m_xL_x = m_y L_y$
implies that they are multiples of $L_xL_y$; namely $m_xL_x = m_y L_y
= p L_xL_y$ with $p$ an integer. Thus we obtain the result:
\be
\nu = \frac{V_{FS}}{(2\pi)^2} + p
\label{luttrel}
\ee
which of course is Luttinger's theorem\cite{Luttinger60} that relates the Fermi volume to
the filling (modulo filled bands that are represented by the integer
$p$). 

\section{Momentum Balance in $FL^*$}
\label{FLstar}
Here we will consider an exotic variant of the Fermi liquid, where
electron like quasiparticles coexist with $Z_2$
fractionalization\cite{Senthil_FLstar}. This state may be obtained beginning with a $Z_2$
fractionalized insulating state of electrons that breaks no lattice
symmetries. We consider a specific model where the spinons
($f^\dag_\sigma$ spin half, charge neutral excitations) are fermionic
and the chargons  ($b^\dag$ spin zero, unit charged excitations) are
bosonic. The electron operator is written as $c^\dag_\sigma=b
f^\dag_\sigma$ and the relevant gauge structure is $Z_2$ which implies
that pairing of spinons is present. If one is deep in the insulating
phase then there is a large gap to the chargons; furthermore, if there
is also a spin gap, then the low energy effective theory is just an
Ising gauge theory. For an insulator with an odd number of electrons per
site, in this regime we may set the chargon number $n_b=0$ and the
spinon number $n_f=1$. The Ising gauge charge at each site is
$(-1)^{n_b+n_f}$, which leads to an {\em odd} Ising gauge theory in
this situation. For an insulator with even number of electrons per
site, an {\em even} Ising gauge theory would result. We now imagine a
situation where the lowest charge carrying excitation in the system is
the electron itself. This could arise if the spinon and chargon form a
tightly bound state so that it has a lower net energy than an isolated
chargon. Doping would then lead to a `Fermi liquid' of electron like
quasiparticles, coexisting with gapped visons, spinons and chargons,
which is the $FL^*$ phase we wish to discuss. It already appears that
a violation of Luttinger's relation may be expected here if we dope an
insulator with an odd number of electrons per unit cell, since only
the doped electrons may be expected to enter the Fermi volume. Here we
will see how momentum balance arguments allows for such a violation,
but nevertheless constrains the possible Fermi surface volumes so that
a generalization of Luttinger's theorem to this exotic class of Fermi
liquids holds.

In order to follow in detail the evolution of the system under flux
threading we study the following model Hamiltonian:
\begin{eqnarray}
H_{FL^*} &=& H^0_e + H^0_{sp-ch}+H_{int}+H_{gauge} \\ 
H^0_e &=& -\sum_{<ij>}t^e_{ij}c^\dag_{i\sigma} c_{j\sigma} \\
H^0_{sp-ch} &=& -\sum_{<ij>}t^c_{ij} \sigma^z_{ij} b^\dag_{i} b_{j} -
\sum_{<ij>}t^s_{ij}\sigma^z_{ij}f^\dag_{i\sigma} f_{j\sigma} \nn\\
&+& \Delta_s \sum_i (f^\dag_{i\uparrow} f^\dag_{i\downarrow} + h.c.) \\
H_{gauge} &=& -K\prod_{\Box} \sigma^z_{ij} + h \sum \sigma^x_{ij}
\label{Hflstar}
\end{eqnarray}
with the constraint on all physical states:
\be
\prod_{j=nn(i)} \sigma^x_{ij} = (-1)^{n_b^i + n_f^i}
\label{constraintflstar}
\ee
and $H_{int}$ denotes the interactions between various which we do not
specify here except for assuming that terms here do not couple to an
externally imposed gauge field that is required for flux
threading. Note, the spinons and chargons are coupled to the $Z_2$
gauge field, while the electrons are of course $Z_2$ gauge neutral. We
have selected, 
for simplicity, an on site pairing interaction for the spinons; while
such onsite pairing terms are absent in microscopic models that forbid
double occupancy of electrons, here we are concerned with universal
aspects of quantum phases which are not affected by this
simplification. 

{\bf Flux Threading in $FL^*$:} We now consider the effect of flux
threading on the ground state of the Hamiltonian (\ref{Hflstar}). In
the presence of a vector potential $A^\uparrow$ ($A^\downarrow$)
coupling to the up (down) spin electrons, the hopping matrix
elements for the up (down) spin electrons is modified to $t^e_{ij}
\rightarrow t^e_{ij}e^{iA^\uparrow_{ij}}$ ($t^e_{ij}
\rightarrow t^e_{ij}e^{iA^\downarrow_{ij}}$) , while the chargon
hopping amlitude is modified to $t^c_{ij} \rightarrow
t^c_{ij}e^{\frac{i}{2}(A^\uparrow_{ij}+A^\downarrow_{ij})}$ and the
up (down) spinon hopping amplitude is modified to $t^s_{ij} \rightarrow
t^s_{ij}e^{\frac{i}{2}(A^\uparrow_{ij}-A^\downarrow_{ij})}$ ($t^s_{ij}
\rightarrow
t^s_{ij}e^{-\frac{i}{2}(A^\uparrow_{ij}-A^\downarrow_{ij})} $). Below
we imagine threading $2\pi$ flux in $A^\uparrow$ and study the
evolution of the ground state of the system in this process. In
addition to the excitation of particle-hole pairs of the electron like
Fermi-liquid quasiparticles, we will also see that in some situations
a vison excitation is inserted through the cylinder which gives rise
to the modified Luttinger relations. We begin by considering flux
threading in the absence of vison dynamics ($h=0$ in
Eqn. \ref{Hflstar}), where results are easily derived, and then
reinstate the vison dnamics and show that the central result is
unaffected.

The adiabatic insertion of a unit flux quantum that couples to the up
spin electrons is affected by introducing a gauge field on the
horizontal links of the cylinder in Figure \ref{cylinder} and
increasing its strength from zero ($A^\uparrow_{ij}=0 \rightarrow
2\pi/L_x$) in time $T$. The time evolution of the quantum state can be
written as $|\psi(T)\rangle = \cU_T|\psi(0)\rangle$ where $\cU_T={\cal T}_t
\exp\left(-i \int_0^T H_{FL^*}(t) dt\right)$ where ${\cal T}_t$ is the
time-ordering operator, and the time dependence of the Hamiltonian
arises from the flux threading. Clearly, since $2\pi$ of flux is
invisible to the electrons, the final state must be some excited state
of the initial ($A^\uparrow=0$) Hamiltonian. In order to make this
explicit, the $2\pi$ flux is gauged away, which can be accomplished by
the operators $\cU_\sigma \cU_\phi$ with 
\bea
\cU_\phi &=& \exp \left\{ i\frac{2\pi}{L_x}\sum_i 
x_i (n^e_{\uparrow i}+\frac12(n^i_{f\uparrow}-n^i_{f\downarrow}+n^i_b))
\right\} \nn \\
\cU_\sigma &=& \prod_{ij\in{\rm cut}} \sigma^x_{ij}
\eea

While the first unitary operator eliminates the gauge field for the
electrons, it changes the sign of the hopping matrix element on a
single column of horizontal links for the chargons and spinons which
behave like half charges. This modification to the hopping can be
absorbed in the $Z_2$ gauge fields, which is accomplished by the
unitary operator $\cU_\sigma$, which returns us to the initial
Hamiltonian.  

The action of the time evolution operator $\cU_T$ is to excite
electron like quasiparticles about the Fermi surface in the usual
manner, while the gapped spinons and chargons are not excited during
this adiabatic flux threading. In the absence of vison dynamics ($h=0$
in equation \ref{Hflstar}), it may be seen that a vison is also
introduced during the flux threading proceedure. This is argued as
follows. In the absence of vison dynamics, the vison number through
the hole of the cylinder, as measured by the operator $W_C=\Pi_C
\sigma^z_{ij}$, where $C$ is a contour that winds around the cylinder,
is a good quantum number since it commutes with the Hamiltonian in this
limit. However, in the course of flux threading and returning to the
original gauge, it changes sign since
it may be easily verified that $\cU_\sigma W_C \cU_\sigma^{-1} =
-W_C$, which implies vison insertion. Thus, the final state has a
displaced Fermi sea and a vison.

We can now combine the results of trivial momentum counting and a
knowledge of the vison momentum to obtain the volume of electron like
quasiparticles. This is most easily done in the limit of a very large
gap to the gauge charged particles (spinons and chargons). Then, the
phase is described by gauge neutral electrons forming a Fermi liquid
and a pure Ising gauge theory in the deconfined phase. We know that
the vison excitations of the latter through the hole of the cylinder
with an odd number of rows, carries crystal momentum $0$ or $\pi$
depending on the even or odd nature of the Ising gauge theory. The
gauge constraint in Eqn. \ref{constraintflstar} tells us this depends
on the parity of $n_f+n_b$ at each site. If we set $n_b=0$ for the
gapped chargons and $n_f=1$ for the gapped spinons, where for the
latter we assume the system is obtained continuously by doping a spin
liquid with spin $1/2$ per unit cell (i.e. a spin version of
$I^*_{odd}$). In this limit an odd Ising gauge theory will be obtained
, where the vison threading an odd width cylinder carries crystal
momentum $\pi$. If the gap to the spinons and chargons is now reduced
from infinity, this crystal momentum assignment to the vison cannot
change continuously (from time reversal symmetry), and is hence
expected to be invariant for a finite range of gap values. Thus, the
phase is expected to be continuously connected to the large gap
situation with integer or half integer filling, which will determine
the momentum assignments. 

Thus, we are left with the result that two types of exotic Fermi
liquid states $FL^*_{even}$ and $FL^*_{odd}$ are expected, that differ
in the crystal momentum carried by the vison excitations. The momentum
balance argument then immediately implies that these two states will
have different Fermi volumes at the same filling --- while $FL^*_{even}$
will have a Fermi volume that is identical to that of a conventional
Fermi liquid at the same filling and hence respects Luttinger's
relation, $FL^*_{odd}$ has a Fermi volume that violates Luttinger's
relation in a  very definite way. 

Since the only situation where the momentum balance argument will give
a result distinct from that of a conventional Fermi liquid is for the
case of $FL^*_{odd}$ on an even$\times$odd lattice, where the vison
carries a nontrivial crystal momentum, we discuss that below. Consider
flux threading in such a phase on a cylinder with an odd number of
rows. This will introduce a vison through the hole of a cylinder which
carries crystal momentum $\pi$. This needs to be subtracted from the
usual momentum balance relations for a Fermi liquid displayed in
Eqn. \ref{balancefl}. This leads to a modified Luttinger relation
between the Fermi volume in $FL^*_{odd}$ and and the electron filling $\nu$:
\be
\{\nu - \frac12\} - p = \frac{V^*_{FS}}{(2\pi)^2}
\ee
where $p$ is an integer that represents filled bands. The crucial
difference from the usual Luttinger relation in Eqn. \ref{luttrel} is
the fact that the Fermi volume is determined by $\nu-\frac12$, which
is related to the fact that it is obtained by doping the
fractionalized spin model which is translationally symmetric at half
filling.

Finally we argue that reintroducing the vison hopping ($h\neq 0$) does
not affect these conclusions. In the cases where the vison threading
the cylinder carries zero crystal momentum, introducing vison hopping
leads to a mixing of the vison and no vison states in a finite
system. This implies that we are no longer guaranteed to have a vison
on flux threading. However, for the case of $FL^*_{odd}$ on a cylinder
with an odd number of rows, where the vison carries crystal momentum
$\pi$, the nontrivial crystal momentum blocks the tunneling of
visons even on a  finite sized system. This implies that flux
threading does indeed introduce a vison which finally leads to the
modified Luttinger relation. We can see that the $\pi$ crystal momentum
carried by the vison cannot be transferred to the only other gapless
excitations in the problem, the electron like quasiparticles, since
they are gauge neutral.

\section{Conventional superfluid $\supf$}
\label{conventionalSF}
Luttinger's theorem was formulated for Fermi liquids, and we have extended
Oshikawa's argument to show how the theorem must be modified to account
for the existence of gapped spin liquid insulators (which may be 
equivalently viewed as fractionalized bosonic insulators) as well as
fractionalized Fermi liquids. In both cases, the presence of topological
order was crucial. Let us next turn to conventional superfluids and
ask: What property of the superfluid phase is captured by the Oshikawa 
argument, and gets fixed by the particle density? While we focus on the
case of bosonic superfluids, we expect our results to also be applicable
to s-wave superconductors with a large gap, so that the resulting
Cooper pairs may be effectively viewed as bosons. Also, we consider the 
case of neutral bosons (no internal electromagnetic gauge field). 
Again, these results could be applied approximately to the case of charged 
superfluids if the penetration depth is sufficiently large.

Conventional superfluids in two or more dimensions ($D \geq 2$) are
Bose condensed at zero temperature, and have a unique ground state 
(on both cylinders and torii). It is clear that the Oshikawa argument must 
then capture some property of the excitations in the superfluid.
A conventional superfluid supports two kinds of excitations:
the gapless linearly dispersing Goldstone mode of the broken symmetry state
("phonon"), and topological defects, namely vortices.
We show below that it is the Berry phase acquired by a vortex on adiabatically
going around a closed loop that is fixed by the particle density $\nu$, 
independent of the strength and nature of interactions between bosons in
the superfluid.

\subsection{Effective Hamiltonian for $\supf$}
\label{sec:effsf}

We may describe a conventional superfluid most conveniently in a rotor
representation for the bosons --- thus $B^\dagger_\br \to
e^{-i\phi_\br}$, $B^\dagger_\br B_\br\to N_\br$
with $[\exp(i\phi_\br),N_{\br'}]= \exp(i\phi_\br) \delta_{\br\br'}$, and the
Hamiltonian for interacting bosons in these variables takes
the form
\be
\hat{H}_A(\supf)= -t_b \sum_{\la i j \ra} \cos(\phi_i-\phi_j +
Q A_{ij}) + V_{\rm int}[n]
\label{Hsf} 
\ee 
where the interactions may be of the general form $V_{\rm int}[n]
= \sum_{\br\br'} U_{\br\br'} N_\br N_{\br'}$.
%ignore the $2\pi$ periodicity of $\varphi$ (and hence ignore
%charge quantization), we can expand the first term to quadratic
%order in the phase difference. Since $\varphi_\br$ is no longer an
%angular variable and $N_\br$ takes all real eigenvalues, the Hamiltonian
%reduces to a collection of harmonic oscillators in momentum space ---
%these describe the Goldstone mode (``phonons'') of the superfluid. 
The Bose condensed superfluid, which results in dimensions $D \geq 2$ 
when $t_b$ is the largest scale in the Hamiltonian,
supports linearly dispersing ``phonons'', which are 
the Goldstone mode of the broken symmetry. Vortices
appear as topological defects in the phase field in the ordered state,
and there is a nonzero gap to creating vortices in the bulk of the
superfluid.

In the above discussion, we have assumed that the phase variable has 
periodic boundary conditions, namely, $\varphi_{\br+L_x}=\varphi_\br$, 
$\varphi_{\br+L_y} =\varphi_\br$. However, on cylinders/torii the 
superfluid has additional excited states corresponding to creating 
vortices through holes of the cylinders/torii. A state with 
$\varphi_{\br+L_x} =\varphi_\br+2\pi m_x$ corresponds to a strength 
$m_x$ vortex through a hole in the cylinder.

\subsection{Flux threading in $\supf$}

Consider $N$ bosons each with charge $Q$ condensed into a conventional 
superfluid ground state on an $L_x\times L_y$ lattice in the form of a 
torus. Trivial momentum counting tells us that threading flux $\Phi_0=hc/Q$
into the cylinder on which the system lives changes the crystal
momentum by $2\pi\nu L_y$, where $\nu$ is the filling and $L_y$ is the
number of rows of the cylinder.  We show below, using a low-energy 
description of the superfluid, that adiabatic flux threading introduces a 
vortex into the hole of the cylinder/torus. We do this two steps. First,
we turn off the boson interactions which allows us to directly construct 
the final state and see that it corresponds to introducing one vortex.
Next, we turn back on the boson interactions and argue that this does not 
affect the state or change the momentum carried by the vortex.

\medskip

\noindent{\bf Threading flux $\Phi_0$ introduces a vortex:}

Let us adiabatically thread flux $hc/Q$ in the $-\hat{y}$
direction for the above system on a cylinder as in Fig.~\ref{cylinder}
such that starting
from the initial eigenstate $\vert \Psi(0)\ra$ in the absence of flux,
the final state reached is $\vert \Psi(T)\ra$. The final state can of
course be written as  $\vert \Psi(T)\ra = \cU_T \vert \Psi(0)\ra$ with
$\cU_T={\cal T}_t \exp\left(-i \int_0^T H_A(\supf,t) dt\right)$ where
${\cal T}_t$ is the time-ordering operator. We can go to the
$A_{ij}=0$ gauge by making a unitary transformation $H_A(\supf,T) \to
\cU_G H_A(\supf,T) \cU_G^{-1} \equiv H_0(\supf)$ (corresponding to zero
vector potential). Here $\cU_G = \cU_\phi$, with
\be
\cU_\phi=\exp(i \frac{2\pi}{L_x} \sum_i x_i \hat{n}_i)
\ee
Thus, the final state in the
$A_{ij}=0$ gauge is $\vert \Psi_f\ra = \cU_\phi \cU_T \vert
\Psi_i \ra$. 

Let us consider the extreme limit achieved by $V_{\rm int} 
\to 0$. In this case, the system is initially in the full Bose condensed 
state $\vert \exp(i\phi_i)=1\ra$, which is unaffected by the time
development operator $\cU_T$, and
the final state after acting with $\cU_\phi$ has $\vert \exp(i\phi_i)
= \exp(i 2\pi x_i)/L_x\ra$, namely it corresponds to having a single
vortex threading the hole of the cylinder. Trivial momentum counting
then tells us that this vortex carries momentum: $P_{\rm vortex} = 
2\pi\nu L_y$.

\medskip

\noindent{\bf Flux threading in the presence of boson interactions:}

In the presence of boson interactions, the phase $\varphi$ at each
site is no longer a $c$-number. Since the number and phase do not
commute, interactions introduce phase fluctuations. Such fluctuations
may permit the vortex to escape from the hole of the cylinder. Clearly,
this is only possible if the initial and final state have the same
crystal momentum quantum number. Since the vortex state carries $P_{\rm vortex} 
= 2\pi\nu L_y$, we expect the vortex will remain trapped except at
special values of $\nu$ and $L_y$, where $P_{\rm vortex}$ becomes
a multiple of $2\pi$. Thus, in general, even in the presence of boson 
interactions, threading flux $\Phi_0$ introduces one vortex,
carrying the above crystal momentum, into the hole of the cylinder.

\noindent{\bf Flux threading and the Berry phase for a vortex:}

It is well-known that moving vortices in a stationary Galilean
invariant superfluid 
experience the so-called "Magnus force", a force which acts transverse to the 
velocity of the vortex.  A superfluid vortex thus behaves as a charged particle
in a magnetic field, the Magnus force being analogous to the transverse
Lorentz force.  In a lattice system, this
``magnetic field'' seen by the vortex is encapsulated through vector 
potentials living on the links of the lattice, and the vortices pick
up a Berry phase $\chi$ (of the Aharonov-Bohm kind) on going around an
elementary plaquette of the lattice. We will now show that  using the 
momentum balance argument fixes this Berry phase to be $\chi = 2\pi\nu$.

Consider a single vortex on an infinite plane. 
Since the vortex sees a ``flux'' $\chi$ per plaquette of the lattice, the 
unit translation operators for the vortex
satisfy: $T_x T_y = T_y T_x \exp(-i\chi)$. Let $\vert K_X, Y\ra$
represent the state with one vortex with x-crystal momentum $K_X$
located at $y=Y$.
When this vortex is translated by one unit along the $+\hat{y}$-direction,
i.e. to $y=Y+1$, it is straightforward to show that the new state
has x-crystal momentum given by $K_X+\chi$.  
For an antivortex, the translation operators satisfy $\bar{T}_x 
\bar{T}_y= \bar{T}_y \bar{T}_x \exp(i\chi)$, and translating the
antivorton along $+\hat{y}$ changes the crystal momentum to
$K_X-\chi$.

\begin{figure}
\begin{center}
\hspace*{0mm}
\vskip4mm
\centerline{\fig{3.0in}{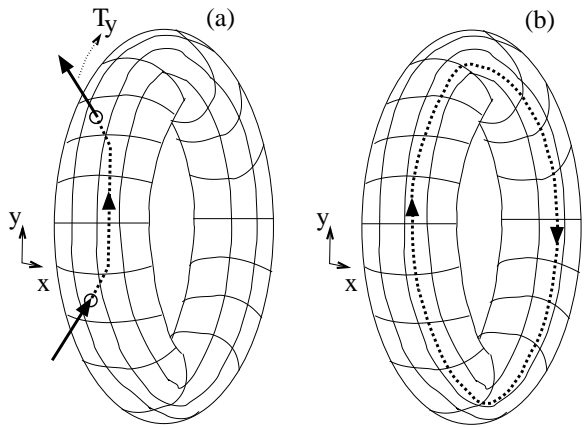}}
\caption{Threading a vortex through the hole of a
torus. (a) A vortex (arrow coming out) - antivortex (arrow
going in) pair is created, and  separated by the translation operator $T_y$.
(b) When the vortex is taken all the way around the torus and
then annihilated with the antivortex, the resulting
state has one unit of circulation about the hole of the torus as shown.}
\vskip-2mm
\label{vortexthreading}
\end{center}
\end{figure}

With this
in mind, let us thread a vortex through the torus in
the manner shown in Fig.~\ref{vortexthreading}.
Start with a state with well-defined crystal
momentum along x-direction, say zero. Create a vortex-antivortex pair
on some plaquette
and make a superposition with {\it zero} net crystal momentum along
$\hat{x}$.  Next drag them apart by translating the vortex 
along $+\hat{y}$ using the translation operator $T_y$
until they are $L$ lattice spacings apart along
the torus. This state then has additional crystal
momentum $ \chi L$. If we drag the pair all the way around the torus
and annihilate the vortex-antivortex pair, this would be equivalent to
threading a vortex through the torus as in Fig.~\ref{vortexthreading}(b). 
The net momentum change is then
$\chi  L_y$. On the other hand, we have shown that
$P_f - P_i = 2\pi\nu L_y$ for threading the vortex through the hole of 
the torus. This fixes $\chi = 2\pi\nu$. 

%%%%%%%%%%%%%%%%%%%%%%%%%%%%%%%%%%%%%%%%%%%%%%%%%%%%%%%%%%%%%%%%%%%%%%
%%%%%%%%%%%%%%%%%%%%%%%%%%%%%%%%%%%%%%%%%%%%%%%%%%%%%%%%%%%%%%%%%%%%%%

We have confirmed this
result by using the well-known duality mapping \cite{U1_duality}
between bosons and
vortices in $2+1$-dimensions. The dual theory treats vortices as 
point particles minimally coupled to a non-compact U(1) gauge field. 
In the dual theory, the flux of the gauge field 
on an elementary plaquette as seen by the vortex
emerges
naturally as $\chi = 2\pi\nu$, i.e. the vortices see each boson as 
a source of $2\pi$ magnetic flux. The non-compactness of the gauge
field, or the conservation of the magnetic flux piercing the
lattice, is a simple consequence of total boson number conservation.

To summarize,
vortices in a uniform superfluid pick up a Berry phase $\chi 
{\cal N}$ on adiabatically going around a loop enclosing ${\cal N}$
plaquettes. The Berry phase per plaquette is completely determined by
the particle density as $\chi=2\pi\nu$. Writing $\chi = 2\pi \alpha_M$,
which defines the ``Magnus coefficient'' $\alpha_M$, 
leads to the Luttinger relation for superfluids, namely
\be
\nu = \alpha_M + p
\label{luttberry}
\ee
This relation follows from
using the momentum balance arguments of Oshikawa, applied to a conventional
superfluid. In this sense, the Berry phase relation above may be viewed
as the analogue of the Luttinger relation for Fermi liquids. 

There is a concern which we have not addressed so far
--- the vortex may have a modified
density near its core, and this in turn could modify 
the Berry phase accumulated by the vortex when it is adiabatically
taken around a loop. However, this term does not change with the
area of the loop, so we can still use the above result to
deduce a precise difference
of the Berry phase between two loops enclosing different areas.
We need to define $2\pi\alpha_M \equiv \Delta\Phi/\Delta{\cal N}$,
where $\Delta\Phi$ is the difference in Berry phase between two loops 
which differ in area by $\Delta{\cal N}$ plaquettes. 
Defined in this manner, $2\pi\alpha_M$ obeys the precise relation in 
(\ref{luttberry}).
Another caveat is the very definition of adiabaticity in the presence of
gapless superfluid phonon excitations --- to be completely rigorous,
we need to work with a finite-sized system such that the phonons
have a nonzero gap, and demand that the
vortex motion be adiabatic with respect to this energy scale.

\section{$Z_2$ fractionalized superfluid $\sfs$}
\label{SFstar}

We now discuss an exotic variant of the superfluid, $\sfs$,
and see how the relation (\ref{luttberry}) is modified in this
phase. The $\sfs$ phase is a $Z_2$ fractionalized superfluid
which was first discussed by Senthil and Fisher.\cite{Senthil_Z2}
It supports three distinct gapped excitations: (i) an elementary $hc/Q$ 
vortex (called the ``vorton'' in Ref.~\onlinecite{Senthil_Z2}),  (ii) $Z_2$
gauge flux 
(the ``vison''), and (iii) an electromagnetically neutral particle 
carrying $Z_2$ gauge charge (the ``ison''). 
There are various ways in which the $\sfs$ phase may be realized, we shall
briefly outline one of them.

Let us start with a fractionalized insulator $\cis$ which can be
realized at integer or half-odd integer density of bosons. This
supports two gapped excitations: charge $Q/2$ chargons that also carry
$Z_2$ gauge charge and Ising
vortices (visons). On doping this insulator, the additional charge-$Q$
bosons can deconfine into pairs of chargons since in the fractionalized phase. Bose
condensing the doped chargons would destroy deconfinement of the $Z_2$
gauge field (by the Anderson-Higgs mechanism, since the condensate 
carries $Z_2$ gauge charge) and
lead to a conventional superfluid.

The other possibility is that doped chargons {\it pair} and Bose
condense resulting in a superfluid phase. Since the condensate is
$Z_2$ gauge neutral, deconfinement is preserved and this exotic
superfluid phase is called $\sfs$. It supports elementary $hc/Q$
vortices, and the visons still survive in the superfluid. There is
however another excitation, analogous to a Bogoliubov quasiparticle of
a superconductor, present in the system --- this gapped 
quasiparticle \cite{Senthil_Z2}
is the ``ison''. It may be viewed as a descendant of the chargon in the
insulator, whose electric charge has been screened by the Bose
condensate of pairs so that it only carries a $Z_2$ Ising charge
\cite{footnote4}. In addition to these gapped excitations, there is of
course the gapless superfluid phonon in an electrically  neutral system.
The relative statistics of the gapped excitations are as
follows: the wavefunction changes sign if an ison is
adiabatically taken around the vison or the vorton.

%The relative statistics of the gapped excitations are as
%follows. Since the ison carries $Z_2$ charge, the wavefunction picks
%up an Aharonov-Bohm phase $\pi$ (i.e., changes sign) if the ison is
%adiabatically taken around the vison.  The ison and the vorton
%have a mutual statistical interaction of the same kind though it is less
%obvious. Roughly speaking, it follows since the chargon-pair picks up
%a phase $2\pi$ going around the vorton, while the ison being one-half 
%of a chargon pair, picks up phase $\pi$.

We have seen how $Z_2$ fractionalized Bose Mott insulators with full lattice
translation symmetry fall into two classes, depending on whether the
boson filling is an integer ($\cise$) or half odd-integer ($\ciso$) as
described in Section \ref{Z2insulator}. Similarly, fractionalized Fermi liquids
$\fls$ also come in two varieties as shown before, with different
relations between fermion filling and Fermi volumes.  It is not
surprising therefore that we will find below two kinds of $Z_2$
fractionalized superfluids, $\sfs$.  Using momentum counting arguments
as done earlier, we will also see how the distinction between the two
types of $\sfs$ phases, namely $\sfse$ and $\sfso$, is reflected in
different dynamics for the vorton in these two cases. While in the case of
translationally symmetric $Z_2$ insulators $\cis$, a knowledge of the
filling alone was enough to determine the odd/even nature of the
phase, this is no longer true in the case of $Z_2$ fractionalized
superfluids, $\sfs$ (or $\fls$), where additional information regarding
the odd/even nature of the phase is needed.

\subsection{Effective Hamiltonian for $\sfs$}
\label{sec:effsfs}

A simple description of the ${\cal SF}^\ast$ phase may be obtained
by using a Hamiltonian which describes chargons minimally coupled
to a $Z_2$ gauge field. This takes the form 
\be
H(\sfs) = \hat{T} + \hat{V} + H_g
\ee
with
\bea
\hat{T}&=& -t_b \sum_{\la i j \ra} \sigma^z_{ij} \cos(\phi_i-\phi_j +
Q A_{ij}/2) \nn\\
&-& t_B \sum_{\la ij \ra} \cos(2\phi_i -2\phi_j + Q A_{ij}) \nn\\
\hat{V}&=& V_{\rm int}[n]  \nn\\
H_g\!\!\! &=&\!\!\! - K \sum_{\Box} \prod_{\Box} \sigma^z_{ij} - 
h \sum_{\la ij\ra} \sigma^x_{ij} \\
\label{Hsfstar}
\eea and physical states of the theory satisfy the constraint in
Eqn. \ref{phys}. Here $\exp(-i\phi_i)$ creates a chargon carrying
charge $Q/2$ and $Z_2$ gauge charge at site $i$, $n_i$ is the
chargon number, and the terms in
$\hat{T}$ represent the chargon kinetic energy. Single-chargons hop
with an amplitude $t_b$, and are coupled minimally to the $Z_2$ gauge
field and for simplicity of discussion, we have included explicitly
a chargon-pair hopping term with amplitude $t_B$. Clearly a chargon-pair
created by $\exp(-2i\phi_i)$ has no net Ising charge and does not
couple to the $Z_2$ gauge field. $V_{\rm int}[n]$ is an interaction
term involving chargon densities which we do not spell out here. The
exotic superfluid phase $\sfs$ requires being in the deconfined phase
of the gauge theory (which is guaranteed by a large $K/h$) and with
chargon pairs condensed (which can be achieved with large chargon pair
hopping $t_B$, while single chargon hopping remains small).

Let us now write down the effective Hamiltonian in the $\sfs$
phase. Condensation of chargon pairs implies that we can replace
the operator $\exp(-2i\phi_i)$ by a c-number. Then, the magnitude of
the chargon creation operator is determined, but its sign can
fluctuate which gives rise to the ison field - i.e. we can write
$\exp(-i\phi_i) \propto I^z_i$, where $I^z_i$ is a Pauli matrix with
eigenvalues $\pm 1$.   
Similarly, since chargon-pairs are condensed, the parity of the
chargon number operator chargon number $n_i$ must be changed by the
ison creation operator $I^z_i$, hence we identify 
$n_i \approx (1+I^x_i)/2$; again $I^x$ is a Pauli matrix and
$(1+I^x)/2$, with eigenvalues $\{0,1\}$, counts the number of unpaired
Ising charged particles. 

In new variables, the Hamiltonian (with $A_{ij}=0$) reduces to
\bea
H_{\rm red} &=& \hat{T}_{\rm red} + \hat{V}_{\rm red} + H_g + 
H_{\rm condensate}\\
\hat{T}_{\rm red} &=& -t'_b \sum_{\la i j \ra} I^z_i I^z_j \sigma^z_{ij} \nn\\
\hat{V}_{\rm red} &=& - g \sum_i I^x_i
\label{Hred_sfstar}
\eea
where we have introduced a ``chemical potential'' $g$ for the isons,
and $H_{\rm condensate} $ describes 
the dynamics of the condensate. Physical states of this
theory need to satisfy the constraint:

\be
 \prod_{j=nn(i)}\sigma^x_{ij}=  -I^x_i
\ee
This is simply an Ising model coupled to an Ising gauge field - the
$\sfs$ phase is realised when both the isons (excitations of the Ising
model) and visons (excitations of the Ising gauge theory) are
gapped. This will occur when $K/h$ and $|g/t'_b|$ are large, when the
gauge theory is in the deconfined phase and the Ising model is
`disordered'. Let us now briefly consider the  special limiting cases
where $|g|\rightarrow \infty$ in order to expose the underlying reason
for the two kinds of $Z_2$ fractionalized $\sfs$ phases.  
Clearly, if $g \to \pm \infty$, we would have $I_i^x=\pm 1$ corrresponding 
to the ison number $(1+I^x_i)/2 = 0,1$ respectively. 
The physical states of the gauge theory then satisfy the constraint:
\be
\underline{g \to \mp \infty:} \hfill \prod_{j=nn(i)} \sigma^x_{ij} = \pm1
\ee

These constraints on the gauge theory, as we know from the discussion
on insulators, correspond to ``even'' and ``odd'' Ising gauge theories
respectively, which correspond to having zero or one Ising charged
particle (ison) fixed at each site. Thus, for $g \to +\infty$ and
for $g \to -\infty$
we will obtain two distinct superfluid phases, which we label $\sfse$ and
$\sfso$ respectively, which persist to finite values of $g$ as well . 
These are
separated by an intermediate phase where $\vert g \vert \ll t_b$ where
the $I^z$ Ising field orders; this is the conventional superfluid
phase. Below, we will see how these $\sfs$ phases can be distinguished
from each other.  

\subsection{Flux threading in $\sfs$} 
Trivial momentum counting tells us that threading flux $\Phi_0=hc/Q$
into the cylinder on which the system lives changes the crystal
momentum by $2\pi\nu L_y$, where $\nu$ is the filling and $L_y$ is the
number of rows of the cylinder. Where is this momentum soaked up in the
$\sfs$ phase?  We will show below that flux threading introduces both
a vison and a vorton into the hole of the cylinder. The crystal
momentum is then divided up between these two excitations, in a way
that depends on whether we are dealing with $\sfse$ or $\sfso$. This
will be argued below in two stages. First, we consider freezing the
Ising gauge field dynamics by setting the vison hopping to zero
($t=0$). There it can easily be argued that $\Phi_0$ flux threading
leads to both a vison and a vorton. Then, using our earlier
knowledge of vison momenta in the even/odd $Z_2$ gauge theories, and
the total momentum imparted to the system, we can read off the crystal
momentum carried by the vorton. Finally, we reinstate the vison
hopping ($t>0$) and use continuity to argue that this does not affect
the momentum assignments.

\medskip

\noindent{\bf Threading flux $\Phi_0$ introduces a vison and vorton:}
Consider at first the limit $h=0$ and $t_B = \infty$ identically, so
that before introducing flux we can everywhere set $\sigma^z_{ij}=1$
(as a reference state which we can then project into the subspace of
physical states) and $\exp(i 2\phi_i) = 1$.

Let us adiabatically thread flux $hc/Q$ in the $-\hat{y}$
direction for the above system on a cylinder as in Fig.~\ref{cylinder}
such that starting
from the initial eigenstate $\vert \Psi(0)\ra$ in the absence of flux,
the final state reached is $\vert \Psi(T)\ra$. The final state can of 
course be written as  $\vert \Psi(T)\ra = \cU_T \vert \Psi(0)\ra$ with 
$\cU_T={\cal T}_t \exp\left(-i \int_0^T H_A(\sfs,t) dt\right)$ where
${\cal T}_t$ is the time-ordering operator. We can go to the
$A_{ij}=0$ gauge by making a unitary transformation $H_A(\sfs,T) \to
\cU_G H_A(\sfs,T) \cU_G^{-1} \equiv H_0(\sfs)$ (corresponding to zero
vector potential). Here $\cU_G = \cU_\phi \cU_\sigma$, with
\bea
\cU_\phi&=&\exp(i \frac{\pi}{L_x} \sum_i x_i \hat{n}_i) \nn\\
\cU_\sigma&=&\prod_{ij\in{\rm cut}} \sigma^x_{ij}
\eea
and 'cut' refers to the vertical column of links for which
$x_i=L_x,x_j=1$ (shown in Fig.~\ref{cut}(b)). Thus, the final state in the
$A_{ij}=0$ gauge is $\vert \Psi_f\ra = \cU_\phi \cU_\sigma \cU_T \vert
\Psi_i \ra$. Since the system is a superfluid initially in the state
$\vert \exp(2i\phi_i)=1\ra$, which is unaffected by the time
development operator $\cU_T$ since we are in the $t_B=\infty$ limit,
the final state after acting with $\cU_\phi$ has $\vert \exp(2i\phi_i)
= \exp(i 2\pi x_i)/L_x\ra$, namely it corresponds to having a single
vorton threading the hole of the cylinder. 

At the same time, following arguments similar to the insulator $\cis$,
acting with $\cU_\sigma$ introduces a vison in the hole, the vison
number being defined by $(1-W_C)/2$ with $W_C=\prod_{C} \sigma^z_{ij}$
where the loop $C$ is taken around the cylinder (see Fig.~\ref{cut}(b)).

Thus, for $h=0, t_B=\infty$, threading a $2\pi$ flux adds a vison and
a vorton into the hole of the cylinder.
Thus, when the effective description of the
gauge fields is an even (odd) Ising gauge theory (which are connected
to the $g \to -\infty$ ($+\infty$) limits respectively as we have seen
above), momentum counting arguments already showed us that the vison
carries momentum zero (for the even gauge theory) or $\pi L_y$ (for the
odd gauge theory). The remaining momentum must clearly be carried by
the vorton!

Thus, in the case of even $Z_2$ gauge theories the vison carries zero
crystal momentum, and we deduce that the vorton through the hole of the
cylinder carries crystal
momentum $2\pi\nu L_y$, and this is the $\sfse$ case.  For the case
of odd $Z_2$ gauge theories , the vison carries crystal momentum $\pi
L_y$ and the vorton carries crystal momentum $2 \pi (\nu-1/2) L_y$,
and this is the $\sfso$ case. To summarize, momentum balance arguments
suggest
\bea
P^{even}_{\rm vorton}&=&2\pi \nu L_y [\mbox{mod } 2\pi] \\
\label{pvorton1}
P^{odd}_{\rm vorton}&=&2\pi (\nu-\frac{1}{2}) L_y [\mbox{mod } 2\pi]
\label{pvorton2}
\eea
We now show that these momentum assignments are
not affected on reinstating the vison hopping ($h
\neq 0)$ and the condensate dynamics ($t_B \neq \infty$).

\medskip

\noindent{\bf Flux threading with dynamical gauge fields:}
Here, starting from the case with $h=0, t_B=\infty$, let us ask what
happens if we turn on a nonzero $h$ and finite $t_B$, giving dynamics 
to the gauge field and to the condensate. 

In this case the loop product $W_C$ no longer commutes with the
Hamiltonian, we cannot use its eigenvalues to label the
eigenstates. Thus, in a finite system as argued previously for the
case of inulators, the vison can tunnel out in the case of even gauge
theories for any cylinder dimension, or in the odd gauge theories, if
the cylinder has an even number of rows ($L_y$). However for the case
of odd gauge theories on a cylinder with odd number of rows ($L_y$),
the vison carried momentum $\pi$, and hence vison tunneling is blocked
even in finite systems. Similarly, when there is dynamics to the
condensate ($t_B \neq \infty$), the vorticity is also not necessarily
conserved in a finite system, namely $\int \nabla 2\phi$ around the
cylinder is not a constant (not a classical variable).  More
precisely, this statement can be made in terms of an order of limits;
if the time scale for flux threading is taken to infinity before the
thermodynamic limit is taken, then the system can remain in the zero
vorticity state at the end of the flux threading.
This occurs if the two states, namely the state with no
vison and no vorton and the state with 1-vison and 1-vorton,
each carry zero crystal momentum, and can then mix to give
eigenstates of the Hamiltonian. This happens if $2\pi\nu L_y = 0
(\mbox{mod }2\pi)$. Otherwise, even in the presence of $h$ and condensate
dynamics, the vorton acquires a nonzero crystal momentum and therefore
its tunneling is blocked even in a finite size system (in the sense
described above). Thus, in all cases where a non trivial crystal
momentum is imparted to the system from flux threading, this is
accounted for by the presence of a vorton and/or a vison in the
final state which carries the appropriate momentum.

The two superfluids $\sfse,\sfso$ can then be distinguished depending
on the momentum carried separately by the vison and the vorton though
the total momentum carried by these excitations is the same. We shall
later see that this may be reflected in the measured Hall effect in
the vorton liquid phase in these systems through the Berry phase
induced Magnus force on the vorton.
In the next subsection, we shall directly verify the momentum
assigments for the vorton by going to a dual
theory where vortices are represented as particles.

\subsection{Berry phase for vorton and consistency with
momentum counting}

The dual theory for the $\sfs$ phase, where the chargons are 
traded for vortex variables, is derived in Appendix A. The action
takes the form

\bea
S&=&S_{gauge} + S_{c} \nn \\
S_{gauge}&=& -\varepsilon K \prod_{\Box_s} \sigma_{ij}
-\varepsilon K_\tau \prod_{\Box_\tau} \sigma_{ij} \nn\\
S_{c}&=& - \varepsilon 
t_v \sum_{i,\mu} \cos(\theta_i -\theta_{i+x_\mu} - 
2\pi\cA^i_\mu - \pi a^i_\mu) \nn \\
&+& - \varepsilon g \sum_i I^i_0 + 
i \frac{\pi}{2} \sum_{i,\mu} I^i_\mu (1-\sigma_{i,i+\tau})  \nn \\
&+& \alpha\sum_{i,\mu=1,2} (\cJ^i_\mu)^2 + \varepsilon U 
\sum_i (\cJ^i_0-\bar{n})^2.
\eea
Here $\exp(-i\theta)$ creates a vorton, and the first term in $S_c$
represents vorton hopping. The vortons appear as charged
particles minimally coupled to the gauge fields $\cA_\mu$ and $a_\mu$.
The flux of the (noncompact) $U(1)$ gauge field $\cA_\mu$ and the
gauge field $a_\mu$ are tied to the charge current ($\cJ_\mu$) and 
the ison current ($I_\mu$) respectively:
\bea
\cJ^i_\mu/2 &=& \epsilon_{\mu\nu\lambda}\partial_\nu \cA^i_\lambda \\
I^i_\mu&=&\epsilon_{\mu\nu\lambda}\partial_\nu a^i_\lambda.
\eea
The isons have a chemical potential $g$, and a Berry phase term
associated with the fact that they couple to the $Z_2$ gauge field
$\sigma_{ij}$. The remaining terms represent local interactions
between the charge density/currents.

On the spatial links ($\mu=1,2$), $\la I^i_\mu \ra = 0$ and
$\la \cJ^i_\mu \ra = 0$.
On the temporal links, for the chargon density we have 
$\la \cJ^i_0\ra = 2\nu$, where $\nu$ is the average charge density
in units of $Q$ (equivalently, $\nu$ is the density of chargon-{\it pairs}). 
For the ison density, we have two possibilities: (i)
for $g \to -\infty$, we have $I^i_0=0$ and there are no isons in
the ground state; (ii) for $g \to +\infty$,
there is one ison nailed down to each lattice site, $I^i_0 = 1$. 
For $g \to -\infty$ (or $\sfse$), when the vortons go anticlockwise around
an elementary plaquette of the square lattice, they see only the flux produced
by the vector potential $2\pi \cA$, and the wavefunction acquires a factor
$\exp(i \pi \cJ^i_0)$. On average, the phase picked up is thus
$2\pi\nu$. This is identical to the Berry phase picked up by a vortex 
in a conventional superfluid with boson density $\nu$.
For $g\to +\infty$ (or $\sfso$), the vorton sees an {\it additional} flux produced by 
one ison charge $I^i_0=1$ nailed down to each site, and the total flux 
seen by the vorton is thus $2\pi (\nu-1/2)$ ---  this deviates by $\pi$
from the Berry phase for $\sfse$ and the conventional superfluid.

Applying the picture of vortex threading presented for conventional
superfluids to this case of vorton threading, it is clear that this
Berry phase is consistent with the momentum counting argument. Namely,
the vorton threading suggests that $\chi L_y = P_{\rm vorton}$.
On the other hand, momentum balance tells us 
$P_f - P_i = 2\pi\nu L_y$ for threading {\it a vison and a vorton}
through the hole of the torus. Since
we know that the vison carries crystal momentum $P^{(even)}_{\rm vison}=0$ 
in an even gauge theory (as in $\sfse$) or $P^{(odd)}_{\rm vison} = \pi L_y$ 
in an odd gauge theory
(as in $\sfso$). This fixes $P_{\rm vorton}$ in the two cases (in agreement
with Eqns.~\ref{pvorton1},\ref{pvorton2}) and thus 
$\chi^{(even)} = 2\pi\nu$ (in $\sfse$),
and $\chi^{(odd)}=2\pi(\nu-1/2)$ (in $\sfso$). This is consistent with the
result derived from the dual theory above. Defining as before the ``Magnus
coefficient'' $\alpha_M = \chi/2\pi$, we obtain
\bea
\alpha^{even}_M &=& \nu - p \\
\alpha^{odd}_M &=& \{\nu-\frac{1}{2}\} - p,
\eea
where $p$ is an arbitrary integer.
Thus the momentum counting
provides a prescription to fix the Berry phase for the vorton, and allows 
us to distinguish the ``odd'' and ``even'' exotic superfluid phases.

\begin{figure}
\begin{center}
\vskip2mm
\hspace*{0mm}
\centerline{\fig{2.5in}{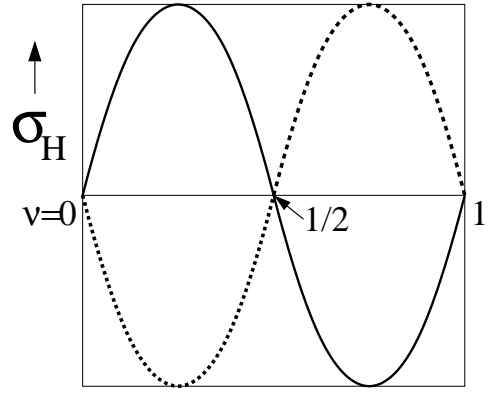}}
\vskip2mm
\caption{Schematic figure showing the filling dependence ($\nu=0$ ($1$) is the empty (full) band) of the
Hall conductivity in a conventional Fermi liquid or $\flse$ (solid
line), for
electrons within a simple Drude-like picture, contrasted
with the behavior expected in the $\flso$ phase (dashed line).
Similar results are expected for the conventional superfluid or
$\sfse$ phase compared to the $\sfso$ phase in a vortex liquid phase.}
\label{hall}
\end{center}
\end{figure}

%A similar situation occurs in superfluids. Let us assume
%that we probe the Hall response of a clean dilute vortex liquid, such that vortex
%interactions are unimportant. The Hall response is then a sum of single vortex
%contributions. In a conventional superfluid or in the $\sfse$, with 
%filling $\nu = 1+\epsilon$ we expect that we
%can translate the Berry phase for the vortex, $2\pi(1+\epsilon)$,
%into a transverse ``Magnus force'' within a semiclassical
%picture. Ignoring $2\pi$ phases, which is possible say for a nearest
%neighbor hopping model for the vortices, the effective Berry phase is 
%just $\chi = 2\pi \epsilon$, which implies a Magnus force $\sim \epsilon$.
%For the $\sfso$
%phase however, the Berry phase at this density is $\chi = 
%(\pi+2\pi\epsilon)$. In simple 
%models of vortex hopping in $\sfso$, this would change the sign of the 
%Magnus force to $\sim -\epsilon$, which for a conventional superfluid would
%be achieved at a filling $-\epsilon$. This translates
%into a different sign for the Hall response of the vortex liquid state 
%in the $\sfso$ phase compared to the $\sfse$ or the conventional 
%superfluid,
%which parallels the difference between a conventional Fermi liquid
%or $\flse$ and the $\flso$ phase (see Fig.~\ref{hall}).
%

\section{conclusions}
Extending a non-perturbative argument, made by Oshikawa for the 
Fermi liquid, we have constructed analogues of Luttinger's theorem
for systems other than the conventional Fermi liquid in dimensions
$D \geq 2$. This has
allowed us to derive constraints which must be satisfied by
quantum phases of matter on a lattice, such as superfluids and the more
exotic $Z_2$ fractionalized phases which are topologically ordered.
We have discussed ways in which these constraints may be useful in 
identifying fractionalized phases in numerical experiments. 

A recurring 
theme has been the important distinction between `even' and `odd'  
deconfined Ising gauge theories, which correspond
to states that are most naturally associated with integer and half 
integer filled systems respectively.
For exotic insulators, the even or odd character of the phase is
completely determined by the filling in this manner. For exotic
Fermi liquids and superfluids, a knowledge of the filling by
itself is insufficient to determine the odd or even nature of the
emergent $Z_2$ gauge field --- precise
violations of the Luttinger relation or its analogue
for these systems provides a way to distinguish them from each other.

Within
a simple Drude-like picture, one associates the size of the
Fermi surface to the sign of the Hall conductivity --- a Fermi 
surface corresponding to a few electrons exhibits an electron-like
response, while a Fermi surface of a nearly filled band would
show a hole-like response.  A Fermi surface which violates the conventional
Luttinger theorem may thus be reflected in an anomalous
sign of the Hall conductivity
as depicted schematically in Fig.~\ref{hall}.
A similar change in the sign of the Hall effect in a vortex
liquid phase is expected
for odd fractionalized superfluids relative to conventional
superfluids, due to a shift of the Berry phase by $\pi$ at 
a given density of bosons.
Clearly, the sign of the Hall effect is not universal and
in real systems is affected by band structure and
interactions. Thus an anomalous sign of the Hall response is
suggestive but is not a rigorous diagnostic.
Experimental
tools such as angle resolved photoemission spectroscopy which
measure the Fermi surface
can more directly detect violations
of the conventional Luttinger theorem expected in odd Fermi
liquids and thus serve to identify such systems.

\section{acknowledgements}
We thank L. Balents, M.P.A. Fisher, 
G. Misguich, M. Oshikawa, and T. Senthil for stimulating discussions. 
AP received support through
grants NSF DMR-9985255 and PHY99-07949 and the Sloan and Packard 
foundations. AV acknowledges support from a Pappalardo Fellowship.

\appendix

\section{Duality and vortons in $\sfs$}

Let us begin with the path integral for the partition function of chargons 
coupled to a $Z_2$ gauge field, $Z = \int \cD\phi \sum_{\{n\}}
\sum_{\{\sigma\}} \exp(-S)$, with 
$S = S_{gauge} + S_c$. The gauge field and chargon actions are given
by
\bea
S_{gauge} &=&  -\varepsilon K_s \prod_{\Box_s} \sigma_{ij}
-\varepsilon K_\tau \prod_{\Box_\tau} \sigma_{ij} \nn\\
S_c&=& -\varepsilon t_b \sum_{\la ij ra} \sigma_{ij} \cos(\phi_i-\phi_j) 
+\varepsilon U \sum_i (n_i-\bar{n})^2 \nn \\
&+& i \sum_i n_i (\phi_i - \phi_{i+\tau} + \frac{\pi}{2}[1-
\sigma_{i,i+\tau}])
\eea
Here, $\la ij\ra$ denotes nearest neighbor sites in space.
$n,\phi$ denote the chargon number and phase. The chargons
hop with an amplitude $t_b$ and have a local repulsion of strength 
$U$. $U\bar{n}$ plays the role of the chargon chemical potential.
The $Z_2$ gauge fields are denoted by $\sigma_{ij}=\pm 1$
and $\Box_s, \Box_\tau$ denote
elementary spatial/spatiotemporal plaquettes on the cubic space-time 
lattice with respective gauge field couplings $K_s,K_\tau$.
Finally, $\varepsilon$ is the Trotter discretization along the 
imaginary time direction.

We can rewrite the chargon hopping term in the partition function
as
\be
\sum_{L_{ij}} e^{-\alpha \sum_{\la ij\ra} L_{ij}^2 + i \sum_{\la ij \ra} 
L_{ij} (\phi_i-\phi_j+\frac{\pi}{2}[1-\sigma_{ij}])}
\ee
where $L_{ij} = - L_{ji}$ is an integer-valued field.
For large $\alpha$, with
$\alpha=\ln(2/\varepsilon t_b)$, this reduces to the original
chargon hopping term.
For general $\alpha$, this modified form allows terms such as
$\cos(2\phi_i - 2\phi_j)$ which correspond to chargon-pair hopping. We
therefore do not need to keep an explicit pair-hopping term $t_B$
unlike in our discussion in section \ref{SFstar}.

Integrating out the phase field $\phi$ leads to a constraint
\be
\sum_j L_{ij} + (n_i - n_{i+\tau}) = 0
\ee
which is just the discrete version of the continuity equation,
$L_{ij}$ representing the chargon current on bond $(i\to j)$.
Writing the
number $n_i \equiv L_{i,i+\tau}$, we can recast this in the form
\be
\sum_{\mu=0,1,2} [L_{i,i+x_\mu} + L_{i,i-x_\mu}] = 0
\ee
where $x_0=\tau$, $x_1=x$, $x_2=y$. This sets the divergence of
the 3-current to zero.  Below, the sum over $\mu$ will
be understood to run over $0,1,2$ unless stated, we will also 
use the notation $L^i_\mu \equiv L_{i,i+x_\mu}$.

We go to dual vortex variables in $2+1$ dimensions in the standard manner 
\cite{U1_duality}, the only difference is in the presence of $Z_2$ gauge fields
in the action but we do not dualize these.
The constraint is solved by equating the conserved current to the
curl of a dual vector such that its divergence is automatically zero.
We decompose the chargon current into
two parts, the current of pairs $J$ (an even integer) and the
current of unpaired particles $I$ ($=0,1$). Note that $I$ is only
conserved modulo-2 --- two unpaired particles can combine to form a pair
which is accommodated by increasing $J$ by one unit, and $I$ thus
is the current of particles carrying only a $Z_2$ charge.
The constraint is thus solved by choosing
\bea
J^i_\mu&=&2 \epsilon^{\mu\nu\lambda} \partial_\nu A^i_\lambda \\
I^i_\mu&=& (\epsilon^{\mu\nu\lambda} \partial_\nu a^i_\lambda) [\mbox{mod } 2],
\eea
where $A$ (an integer) and $a$ ($=0,1$) are
fields on links of the dual space-time lattice, and the right
hand sides above are just the lattice curls on the dual lattice,
taken around the original link $(i,i+x_\mu)$.

The chargon action then takes the form
\bea
S_c&=&  \alpha \sum_{i,\mu=1,2}
(J^i_\mu + I^i_\mu)^2 + \varepsilon U (J^i_0+I^i_0-\bar{n})^2 \\
&+& i \frac{\pi}{2} \sum_i I^i_0 
(1-\sigma_{i,i+x_\mu})\\
&+& g \sum (\epsilon^{0\nu\lambda} \partial_\nu a_\lambda)[\mbox{mod }2]
\eea
where we have now included a chemical potential $g$ for the $Z_2$ 
charges whose density is $I^i_0 = 
\epsilon^{0\nu\lambda} \partial_\nu a_\lambda [\mbox{mod }2]$.

We can convert the sum over $A$ to an integral by softening the constraint
by introducing terms $\varepsilon t_v \sum_{i,\mu} 
\cos(2\pi A^i_\mu)$ in the action (this can be formally accomplished by 
using Poisson resummation), which prefers $A^i_\mu$ to be an integer.
Everywhere else in the action, only the transverse part of $A$ plays
a role (since only its lattice-curl appears).
Extracting the longitudinal part of $2\pi 
A_{ij}$ as $\theta_i-\theta_j$, we identify the dual vorton creation
operator $\exp(-i\theta_i)$. The vortons are seen to be minimally coupled to 
the transverse
part of the $A$, which we denote $\cA$, exactly as a charged particle
coupled to a $U(1)$ gauge field. Thus,
in the softened theory, 
\be
2\pi A^i_\mu = (\theta_{i+x_\mu}-\theta_i) + 2\pi\cA^i_\mu.
\ee
Making this substitution, and absorbing $a$ by shifting $\cA^i_\mu \to 
\cA^i_\mu- a^i_\mu/2$ find
\bea
S_c&=&  \alpha\sum_{i,\mu=1,2} (\cJ^i_\mu)^2 + \sum_i (\cJ^i_0-\bar{n})^2 \\
&+& i \frac{\pi}{2} \sum_{i,\mu} I^i_\mu (1-\sigma_{i,i+\tau}) + 
g \sum_i I^i_0 \\
&-& t_v \sum_{i,\mu} 
\cos(\theta_i -\theta_{i+x_\mu} - 2\pi\cA^i_\mu - \pi a^i_\mu)
\eea
with the total current $\cJ^i_\mu = J^i_\mu + I^i_\mu \equiv 
2 \epsilon_{\mu\nu\lambda}\partial_\nu \cA^i_\lambda$. This is the result 
used in Section \ref{SFstar}~C.


\begin{thebibliography}{999}

\bibitem{Oshikawa97} 
M. Yamanaka, M. Oshikawa and I. Affleck, Phys. Rev. 
Lett. {\bf 79}, 1110 (1997).

\bibitem{Oshikawa00} 
M. Oshikawa, Phys. Rev. Lett. {\bf 84}, 3370 (2000).

\bibitem{Luttinger60}
J.M. Luttinger, Phys. Rev. {\bf 119}, 1153 (1960).

\bibitem{Misguich_numerics}
G. Misguich, C. Lhuillier, B. Bernu, and C. Waldtmann
Phys. Rev. B {\bf 60}, 1064 (1999);
G. Misguich, C. Lhuillier, M. Mambrini and P. Sindzingre,
Eur. Phys. J. B {\bf 26}, 167 (2002).

\bibitem{Sheng04}
D. N. Sheng and L. Balents (unpublished).

\bibitem{Senthil_FLstar}
T. Senthil, S. Sachdev and M. Vojta, Phys. Rev. Lett. {\bf 90}, 216403 (2003).

\bibitem{HaldaneWu85}
F.D.M. Haldane and Y. Wu, Phys. Rev. Lett. {\bf 55}, 2887 (1985).

\bibitem{Ao93}
P. Ao and D.J. Thouless, Phys. Rev. Lett. {\bf 70}, 2158 (1993).

\bibitem{Senthil_Z2}
T. Senthil and M. P. A. Fisher, Phys. Rev. B {\bf 62}, 7850 (2000).

\bibitem{paramekanti98}
A. Paramekanti, N. Trivedi and M. Randeria, Phys. Rev. B {\bf 57}, 11639 (1998).

\bibitem{Moessner02}
R. Moessner, S. L. Sondhi, and E. Fradkin, Phys. Rev. B {\bf 65}, 024504 (2002).

\bibitem{Laughlin81}
R. B. Laughlin, Phys. Rev. B {\bf 23}, 5632 (1981).

\bibitem{Wexler97}
C. Wexler, Phys. Rev. Lett. {\bf 79}, 1321 (1997).

\bibitem{Oshikawa03}
M. Oshikawa, Phys. Rev. Lett. {\bf 90}, 236401 (2003) and Phys. Rev. Lett.
{\bf 91} 109901 (Errata) (2003).

\bibitem{Hastings}
M.B. Hastings, Phys. Rev. B {\bf 69}, 104431 (2004).

\bibitem{controversy}
See D. J. Thouless, P. Ao and Q. Niu, Phys. Rev. Lett. {\bf 76}, 3758 (1996),
G. E. Volovik, Phys. Rev. Lett. {\bf 77}, 4687 (1996),
E. B. Sonin, Phys. Rev Lett. {\bf 81}, 4276 (1998), C. Wexler,
D.J. Thouless, P. Ao and Q. Niu, Phys. Rev. Lett. {\bf 81}, 4277 (1998).

\bibitem{footnote.drude}
Note that we {\it define} an insulator as $\bar{I}=0$ which is a stronger
condition than a vanishing Drude weight used by 
Oshikawa\cite{Oshikawa03}. In particular, a system with a finite zero
frequency conductivity would be an insulator by his definition
but not by the more stringent condition used here.

\bibitem{Oshikawa00b}
M. Oshikawa, Phys. Rev. Lett. {\bf 84}, 1535 (2000).

\bibitem{Shankar}
D. H. Lee and R. Shankar, Phys. Rev. Lett. {\bf 65}, 1490 (1990).

\bibitem{footnote1}
The total number of such quasi-degenerate states depends on the 
pattern of symmetry breaking in the thermodynamic limit. However,
all these states may not be accessible from the ground state by
flux threading; the momentum counting argument provides a
constraint on which states can be accessed. For instance, in
$D=2$, only states which differ from the ground state by 
crystal momentum $2\pi (p/q) L_y$ can be accessed when threading
flux along the $y$-direction.

%If the system is on a cylinder and the
%broken symmetry pattern dictates a common
%factor between $L_y$ (length along the cylinder axis) and $q$, then
%the number of degenerate states in this geometry is atleast
%$q/HCF(L_y,q)$
%where $HCF$ denotes the highest (largest) common factor. If the
%symmetry breaking is purely along the $L_x$ direction, then $L_y$ can
%be arbitrary and there will be atleast 
%$q$ degenerate states. In the opposite
%limit, if the symmetry breaking is only along the $\hat{y}$ direction,
%there will be just one ground state on the cylinder.  Thus the ground
%state degeneracy can in general depend on the cylinder dimensions (we
%suspect it will always be $q$ on a torus).

\bibitem{footnote2}
However, they can still cross an even number of times at other
intermediate flux values which, from time-reversal symmetry, are 
symmetric about $\Phi=\pi$.

\bibitem{Microscopicmodels}
R. Moessner and S. L. Sondhi, Phys. Rev. Lett. {\bf 86} 1881 (2001);
L. Balents, M.P.A. Fisher and S.M. Girvin,
Phys. Rev. B {\bf 65}, 224412 (2002);
G. Misguich, D. Serban and V. Pasquier, Phys. Rev. Lett. {\bf 89}, 137202
(2002);
O. Motrunich and T. Senthil, Phys. Rev. Lett. {\bf 89}, 277004 (2002).
A. Kitaev, Ann. Phys. {\bf 303}, 2 (2003);
X.-G. Wen, Phys. Rev. Lett. {\bf 90}, 016803 (2003).

\bibitem{footnote3}
This is to be viewed as choice of a
reference state which we can then project into the subspace of
physical states which satisfy the constraint (\ref{phys}).

\bibitem{Bonesteel89}
N.E. Bonesteel, Phys. Rev. B {\bf 40}, 8954 (1989).

\bibitem{Anderson87}
P. W. Anderson, Science {\bf 235}, 1196 (1987).

\bibitem{wavefunctions}
D. Ivanov and T. Senthil, Phys. Rev. B {\bf 66}, 11511; A. Paramekanti,
M. Randeria and N. Trivedi (cond-mat/0303360,0405353).

\bibitem{footnote.mse}
We have also checked the vison momentum to be expected on a triangular
lattice on even$\times$even and odd$\times$even geometries. On
even$\times$even lattices, it is still zero, while on odd$\times$even
lattices it carries a momentum which is half of the reciprocal lattice
vector.

\bibitem{U1_duality}
M. P. A. Fisher and D. H. Lee, Phys. Rev. B {\bf 39}, 2756 (1989).

\bibitem{footnote4}
A different way to view ${\cal SF}^\ast$ 
is to consider a weakly coupled bilayer system consisting of two parts:
(i) a fractionalized insulating
layer $\cis$ which supports gapped visons
and chargons and (ii) a condensate of charge-$Q$ bosons
which is a conventional superfluid
(${\cal SF}$) supporting $hc/Q$ vortices. In the absence
of any coupling between the layers, this bilayer system clearly
supports all these separate excitations. Imagine turning on a weak
coupling which allows transfer of charge between the two layers, via
processes which involves two chargons in the $\cis$-layer
pairing and becoming part of the condensate in the superfluid
layer. The chargon is then no longer a well-defined excitation in the
system since its electric charge can be screened by the
condensate. The physical excitation is in fact an electrically neutral
remnant of a single chargon, which carries only a $Z_2$ Ising gauge
charge, the ison. The vison and the $hc/Q$ vortex
continue to be good excitations.

\end{thebibliography}
\end{document}